\documentclass[preprint,12pt]{elsarticle}

\usepackage{preambles}

\newcommand{\MDDataStructure}{Lattice List}
\newcommand{\MDDataStructureLowerCase}{lattice list}
\newcommand{\MDOffset}{Neighbor Offset Index}
\newcommand{\MDOffsetLowerCase}{neighbor offset index}

\def\BibTeX{{\rm B\kern-.05em{\sc i\kern-.025em b}\kern-.08em
    T\kern-.1667em\lower.7ex\hbox{E}\kern-.125emX}}

\newcommand{\new}[1]{#1}
\newcommand{\add}[1]{{\color{black}{#1}}}
\newcommand{\change}[1]{{\color{black}{#1}}}

\newcounter{bla}

\journal{Computer Physics Communications}

\begin{document}

\begin{frontmatter}




\title{MD Simulation of Hundred-Billion-Metal-Atom Cascade Collision on Sunway Taihulight}

\maketitle


\author[ustb,ercis]{Genshen Chu\corref{author}}
\author[ustb,ercis]{Yang Li\corref{cor_author}}
\author[ustb]{Runchu Zhao}
\author[ustb]{Shuai Ren}
\author[ciae]{Wen Yang}
\author[ciae]{Xinfu He}
\author[ercis]{Chungjun Hu}
\author[ercis]{Jue Wang}

\cortext[author] {\textit{E-mail address:} genshenchu@gmail.com}
\cortext[cor_author] {Corresponding author.\\\textit{E-mail address:} liyang@ustb.edu.cn}
\address[ercis]{Engineering Research Center of Intelligent Supercomputing,Ministry of Education, Beijing, China}
\address[ustb]{University of Science and Technology Beijing, Beijing, China}
\address[ciae]{China Institute of Atomic Energy, Beijing, China}
\address[sccas]{Computer Network Information Center, Chinese Academy of Science, Beijing, China}

\begin{abstract}
    Radiation damage to the steel material of reactor pressure vessels is a major threat to the nuclear reactor safety.
    It is caused by the metal atom cascade collision, initialized when the atoms are struck by a high-energy neutron.
    The paper presents MISA-MD, a new implementation of molecular dynamics, to simulate such cascade collision with EAM potential.
    MISA-MD realizes (1) a hash-based data structure to efficiently store an atom and find its neighbors,
    and (2) several acceleration and optimization strategies based on SW26010 processor of Sunway Taihulight supercomputer,
    including an efficient potential table storage and interpolation method,
    a coloring method to avoid write conflicts, and double-buffer and data reuse strategies.
    The experimental results demonstrated that MISA-MD has good accuracy and scalability,
    and obtains a parallel efficiency of over 79\% in an 655-billion-atom system.
    Compared with a state-of-the-art MD program LAMMPS,
    MISA-MD requires less memory usage and achieves better computational performance.
\end{abstract}


\end{frontmatter}



{\bf PROGRAM SUMMARY/NEW VERSION PROGRAM SUMMARY}

\begin{small}
\noindent
{\em Program Title: } MISA-MD \\
{\em CPC Library link to program files:} (to be added by Technical Editor) \\
{\em Developer's repository link:} (if available) \\
{\em Code Ocean capsule:} 10.24433/CO.4041607.v1  \\
{\em Licensing provisions(please choose one):} BSD 3-clause  \\
{\em Programming language: } C and C++ \\
{\em Supplementary material:} \\
{\em Journal reference of previous version: }*                  \\
{\em Does the new version supersede the previous version?:}*   \\
{\em Reasons for the new version:*}\\
{\em Summary of revisions:}*\\
{\em Nature of problem(approx. 50-250 words):}\\
  Molecular dynamics(MD) is \change{a} significant method to simulate
  the cascade collision progress of the key material in nuclear \change{reactors}.
  However, there are many difficulties for existing MD programs to perform large scale cascade collision \change{simulations}.
  Thus, it is especially essential to develop a new MD software to extend cascade collision \change{simulations} 
  to larger spatial scale and longer temporal scale. \\
{\em Solution method(approx. 50-250 words): } \\
  To achieve accuracy and effective MD cascade collision simulation,
  \add{the} EAM potential is selected to calculate interactional force between atoms in \add{the} simulation system.
  To extend MD simulation to larger scale,
  we proposed a hash-based data structure/algorithm to efficiently store an atom and find its neighbors,
  and several acceleration and optimization strategies based on SW26010 processor of Sunway Taihulight supercomputer. \\
{\em Additional comments including restrictions and unusual features (approx. 50-250 words):}\\
   \\

* Items marked with an asterisk are only required for new versions
of programs previously published in the CPC Program Library.\\
\end{small}


\section{Introduction}
As a sustainable, clean and renewable energy source, nuclear energy is a promising solution to global energy crisis and environmental pollution.
Nevertheless, how to ensure the safety of the nuclear reactor is a crucial issue.
Many components of a reactor are exposed to a radiation environment that is of high temperature and high pressure, and is full of high-energy neutrons.
Particularly, reactor pressure vessel (RPV), a unique and non-replaceable part, is the last protection of the fission reactor core.
Its integrity directly determines the service time of the whole reactor.

Radiation damage to the key material of RPV is a major threat to the integrity of RPV.
Existing research revealed that the damage is initiated when a given lattice atom, namely, primary knock-on atom (PKA), is struck by a high-energy neutron \cite{stoller_primary_2012,knaster_materials_2016,nordlund_primary_2018,nordlund_historical_2019}.
Then, PKA will continue to perform a sequence of collisions with other atoms.
Afterwards, the system will generate the secondary, the third, and the subsequent higher-order knock-ons until all the energy initially imported to the PKA has been dissipated \cite{stoller_primary_2012,knaster_materials_2016}.
This process, called \textit{cascade collision}, usually stops within tens of picosecond ($1 \si{\ps} = 10^{-12}\,\si{\second}$) approximately.
It is almost impossible to observe in experiment.


To study material evolution behavior at atomic scale, 
molecular dynamics (MD) \cite{peng_shockwave_2018,fu_molecular_2019} can be applied to the simulation of the \textit{cascade collision}.
Due to the high \change{computational} cost required by the simulation on large spatial scale and long temporal scale, parallel molecular dynamics simulation is especially essential.
There have been a number of parallel molecular dynamics programs \cite{plimpton1993fast, duan_redesigning_2018, niethammer_ls1_2014}.
Nevertheless, existing programs are limited in the functionality 
\add{(e.g. direct defect analyzing, adaptive timestep length, batch execution of simulations)} and the execution efficiency.
Consequently, they are not applicable to the large-scale \textit{cascade collision} simulation.

This paper proposes MISA-MD, a new parallel MD implementation, to achieve the efficient and accurate \emph{cascade collision} simulation 
on large spatial scale and long temporal scale. 
\change{There are various potential functions used in MD simulation under diffrent fields,
such as Tersoff potential \cite{tersoff_new_1988} and Lennard-Jones (L-J) potential \cite{lennard1931cohesion}, for calculating the interaction among atoms.
To improve the simulation accuracy, MISA-MD adopted Embedded Atom Method (EAM) potential \cite{daw_embedded-atom_1984}, 
a complex but pretty accurate potential function, which can provide an effective interatomic description for metallic system.
}To improve the runtime performance, MISA-MD designed and realized a new hash based data structure for efficient atom storage and quick neighbor atom indexing.
Several acceleration and optimization strategies were also applied to ensure that MISA-MD can make full use of SW26010 processors on Sunway Taihulight supercomputer.

The key contributions of this paper are summarized as follows:
\begin{enumerate}
    \item A new hash based data structure for storing and indexing lattice-based atoms;
    \item A new method of EAM potential tables storage and interpolation on SW26010 processor;
    \item A new coloring method to avoid the write conflicts on computing processing elements (CPEs) of SW26010 processor.
    \item New double-buffer DMA and data reuse strategies to reduce the overhead of data transmission from main memory to local device memory (LDM) of CPEs.
\end{enumerate}

The remainder of this paper is organized as follows.
\cref{sec:background} presents the background knowledge on MD and Sunway Taihulight supercomputer, and discusses the related work.
\cref{sec:data-structure} introduces the new atom storage and neighbors indexing data structure.
\cref{sec:sunway} discusses the implementation on Sunway TaihuLight supercomputer and the optimization strategies.
\cref{sec:results} presents a performance analysis and validation of MISA-MD.
The last section summarizes the conclusion and the future work.


\section{Background} \label{sec:background}

\subsection{Molecular Dynamics Method and Challenges}
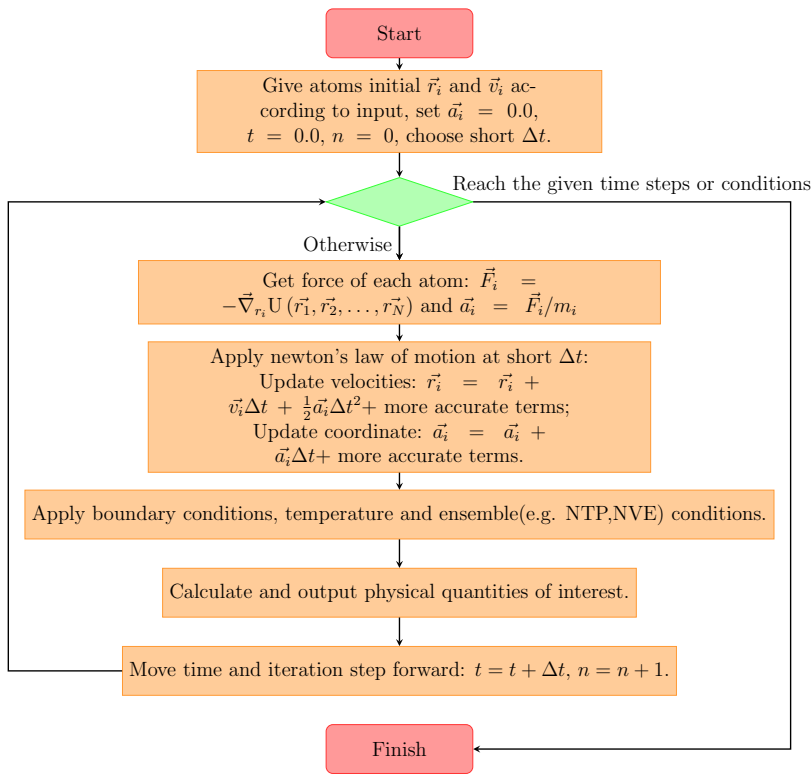
\begin{figure}[!t]
    \centering
    \begin{adjustbox}{scale=0.65,center}
         

\tikzstyle{startstop} = [rectangle, rounded corners, minimum width=3cm, minimum height=1cm,text centered, draw=red!80, fill=red!40]
\tikzstyle{io} = [trapezium, trapezium left angle=70, trapezium right angle=110, minimum width=3cm, minimum height=1cm, text centered, draw=blue!80, fill=blue!30]
\tikzstyle{more} = [ellipse, minimum width=4cm, minimum height=1.5cm, text centered, draw=blue!80, fill=blue!40]
\tikzstyle{process} = [rectangle, minimum width=3cm, minimum height=1cm, text centered, draw=orange!80, fill=orange!40]
\tikzstyle{decision} = [diamond, aspect=6, minimum width=3cm, minimum height=1cm, text centered, draw=green!80, fill=green!30]
\tikzstyle{arrow} = [thick,->,>=stealth]

\begin{tikzpicture}[node distance=1.6cm]
    \node (start) [startstop] {Start};
    \node (init_particles) [process, below of=start, text width=8cm] {
        Give atoms initial $\vec{r}_{i} $ and $\vec{v}_{i}$ according to input,
        set $\vec{a_i}=0.0$, $t=0.0$, $n=0$, choose short $\Delta t$.
    };

    \node (step_for) [decision, below of=init_particles, yshift=-0.25cm] {
    };

    \node (get_force) [process, below of=step_for, yshift=-0.25cm, text width=10.4cm] {
        Get force of each atom: $\vec{F_i} = -\vec{\nabla}_{r_i} \mathrm{U}\left(\vec{r_1}, \vec{r_2}, \ldots, \vec{r_N} \right)$
        and $\vec{a_i}=\vec{F_i} / m_i $
    };

    \node (update_position) [process, below of=get_force, yshift=-0.75cm, text width=10cm] {
        Apply newton's law of motion at short $\Delta t$: \\
        Update velocities:
        $\vec{r_i} = \vec{r_i} + \vec{v_i} \Delta t + \frac{1}{2} \vec{a_i} \Delta t^2 + $ more accurate terms; \\
        Update coordinate:
        $\vec{a_i} = \vec{a_i} + \vec{a_i} \Delta t + $ more accurate terms.
    };

    \node (conditions) [process, below of=update_position, yshift=-0.6cm] {
        Apply boundary conditions, temperature and ensemble(e.g. NTP,NVE) conditions.
    };

    \node (output) [process, below of=conditions] {
        Calculate and output physical quantities of interest.
    };

    \node (next_step) [process, below of=output] {
        Move time and iteration step forward: $t = t + \Delta t$, $n = n + 1$.
    };

    \node (finish) [startstop, below of=next_step] {Finish};

    \draw [arrow] (start) -- (init_particles);
    \draw [arrow] (init_particles) -- (step_for);

    \draw [arrow] (step_for) -- node[anchor=east] {Otherwise} (get_force);
    \draw [arrow] (step_for) -- ++(8.0,0) node[above,midway] {Reach the given time steps or conditions} |- (finish);
    \draw [arrow] (next_step) -- ++(-8.0,0) |- (step_for); 

    \draw [arrow] (step_for) -- (get_force);
    \draw [arrow] (get_force) -- (update_position);
    \draw [arrow] (update_position) -- (conditions);
    
    \draw [arrow] (conditions) -- (output);
    \draw [arrow] (output) -- (next_step);

\end{tikzpicture}
    \end{adjustbox}
    \caption{A standard workflow of MD applications.
    In the time step loop, it will repeat to calculate force, 
    update velocities and position and perform constraint condition as they need.}
    \label{fig:md_algorithm}
\end{figure}

Molecular dynamics is a classic method for simulating particle systems.
It has been widely used in many domains, such as material science, chemistry, and biomolecular science \cite{rapaport_art_2004}.
A generic MD workflow is shown in \cref{fig:md_algorithm}.
In MD, an atom/molecule is treated as a particle.
For initialization, each particle is created with an initial coordinate and a velocity.
Then, the computation falls into a loop of time steps to solve Newton's equations of motion.
In each time step, the computation updates every particle as follows.
First, calculate the force applied to each particle to solve its acceleration $a$.
The force can be calculated by interactional potential functions (e.g. Lennard-Jones potential and EAM potential) in a particular system.
We will discuss the potential function in \cref{sec:eam-pot} in details.
Second, update the particle velocity via the integral of acceleration in a short $\Delta t$,
where $\Delta t$ is the length of a time step.
Third, update the particle coordinate using the integral of velocity in $\Delta t$.
Finally, the corresponding ensemble, \change{such as NVT or NPT ensemble}, is applied, 
and some physical quantities may also be calculated before the loop moves into the next time step.

The interaction forces to be computed among particles can typically fall into two categories:
long-range and short-range \cite{law_algorithm_2019}.
Long-range force calculation \change{requires interactions of each particle contributed by the global system}.
While in short-range force calculation, for any particle $p$,
only its neighbor particles that are located within a cut-off radius is considered to contribute to the force that applied to $p$.
In fact, short-range force calculation is \change{adopted} by most particle systems in MD simulation,
including particle interaction modeled by EAM or L-J potential.

Although most MD implementations are parallel, 
it is still challenging to realize the simulation of large spatial scale and long temporal scale.
For spatial scale, the simulation is mainly restricted by memory storage.
Given a fixed memory capacity, the more particles can be stored (i.e, the less memory is required by a single particle),
the larger spatial scale MD simulation can reach.
For temporal scale, the faster the computation of each time step is, 
the more time steps can be computed within the same time limitation.
For MD simulation with computation of short-range interaction forces,
the time cost of each time step is mainly affected by neighbor particle indexing and the force calculation.
To reduce the time cost of each time step, we must manage the particle data 
and index neighbor particles as efficiently as possible.

\subsection{EAM Potential} \label{sec:eam-pot}

Different from pair potentials (e.g. Lennard-Jones and Morse), embedded-atom method potential is able to provide a better interatomic description for metallic system.
In EAM, the energy of an atom $i$ is determined by two aspects \cite{daw_embedded-atom_1984}:
(1) the embedded energy of the atom embed in electron cloud which represents the \emph{many-body effects} in the interaction,
and (2) the pairwise potential of atomic interaction.
EAM potential can be more complex compared with pair potentials.
In \cref{eq:eam_base}, $E_i$ denotes the energy of atom $i$,
and $r_{ij}$ represents the distance between atom $i$ and its neighbor atom $j$.
Three kinds of functions are presented in \cref{eq:eam_base}:
\begin{enumerate}
    \item $\rho_{\beta}(r)$: It describes the contribution to the electronic density
    at the site with distance of $r$ from atom $j$ whose type is $\beta$.
    \item $F_{\alpha}(\rho)$: The embedding energy function $F$ returns
    the energy associated with placing an atom of type $\alpha$ in the electron environment described by $\rho$.
    \item $\phi_{\alpha \beta }(r_{ij})$: $\phi_{\alpha \beta }$ is pair-wise potential function, which describes
    the pair potential between two atoms $i$ of type $\alpha$ and $j$ of type $\beta$.
    And $r_{ij}$ denotes the distance of atom $i$ and $j$.
\end{enumerate}

\begin{equation}\label{eq:eam_base}
    E_{i}=F_{\alpha}\left(\sum_{j\neq i} \rho_{\beta}(r_{ij})\right) + {\frac {1}{2}}\sum_{i\neq j}\phi _{\alpha \beta }(r_{ij})
\end{equation}

The total force $\vec{F_i}$ applied to atom $i$ can be expressed by
\change{
the negative gradient of potential energy of the simulation system, as descripted in \cite{plimpton_parallel_1992, zhigilei2014introduction}.
Then, we can calculate the force that is applied to atom $i$ from neighbor atom $j$ using \cref{eq:eam_force_ij},
}
where:
    $F_i(\rho)$ and $\rho_i(r)$ denote the embedding energy function and electronic density function 
    associated with type of atom $i$ respectively;
    $\phi_{ij}(r)$ denotes the pairwise potential function associated with types of atom $i$ and $j$;
    $\vec{r}_i$ and $\vec{r}_j$ represent the position vector of atom $i$ and $j$, respectively, and $r_{ij}$ is the distance of atom $i$ and $j$;
    $\bar\rho_i$ denotes the electronic density contributed by all neighbor atoms of atom $i$. 

\begin{figure*}[!t]
    \normalsize
    \begin{equation} \label{eq:eam_force_ij}
        \begin{split}
        \vec{F}_{ij}  = - \left[ 
            \left.\frac{\partial F_i (\rho) }{\partial \rho}\right|_{\rho=\bar\rho_i }
            \left.\frac{\partial \rho_j (r) }{\partial r} \right|_{ r =r_{ij} } +
            \left.\frac{\partial F_j (\rho) }{\partial \rho}\right|_{\rho=\bar\rho_j }
            \left.\frac{\partial \rho_i (r) }{\partial r} \right|_{ r =r_{ij} } +
            \left.\frac{\partial \phi_{ij}(r) }{\partial r}\right|_{ r=r_{ij} }
            \right]
            \frac{ \vec{r}_i -\vec{r}_j }{ r_{ij} }
        \end{split}
    \end{equation}
\end{figure*}

In a system with $n$ atom types, $n \times \left(n + 1\right)/2 + n + n + n$ functions should be determined,
including 
$n \times \left(n + 1\right)/2$ partial derivative of pairwise potential functions,
$n$ partial derivative of embedding energy functions,
$n$ electronic density functions and $n$ partial derivative of electronic density functions.
For example, a system with two different atom types (i.e., $\alpha$ and $\beta$), 
we have to calculate following 9 functions:
$\rho_{\alpha}(r)$, $\rho_{\beta}(r)$, 
$\dfrac{\partial \phi_{\alpha \alpha}(r)}{\partial r}$,
$\dfrac{\partial \phi_{\alpha \beta}(r)}{\partial r}$,
$\dfrac{\partial \phi_{\beta \beta}(r)}{\partial r}$,
$\dfrac{\partial F_{\alpha}(\rho)}{\partial \rho}$,
$\dfrac{\partial F_{\beta}(\rho)}{\partial \rho}$,
$\dfrac{\partial \rho_{\alpha} (r)}{\partial r}$,
$\dfrac{\partial \rho_{\beta} (r)}{\partial r}$.

To calculate the force of each atom using EAM potential according to \cref{eq:eam_force_ij},
the following steps must be performed:
\begin{enumerate} \label{sec:eam_computation_steps}
    \item Calculate electronic density contribution from neighbor atoms for each atom $i$:
    $\bar \rho_i = \sum_{k\neq i} \rho_{k}\left(r_{ik}\right)$;
    \item Calculate value of partial derivative of embedding energy function for each atom $i$ at $\bar \rho_i$:
    $ \left.\frac{\partial F_i (\rho) }{\partial \rho}\right|_{\rho=\bar\rho_i } $;
    \item Calculate value of partial derivative of pairwise potential function for every two atoms $i$ and $j$ in cut-off radius:
    $ \left.\frac{\partial \phi_{ij}(r) }{\partial r}\right|_{r =r_{ij} } $;
    \item For every two atoms $i$ and $j$ in cut-off radius,
    calculate each other's contribution value of partial derivative of electronic density function via:
    $ \left.\frac{\partial \rho_i (r) }{\partial r} \right|_{ r =r_{ij} } $ and
    $ \left.\frac{\partial \rho_j (r) }{\partial r} \right|_{ r =r_{ij} } $;
    \item Here, all terms in \cref{eq:eam_force_ij} are avaiable,
    then we can calculate force $\vec F_{ij}$ of two atoms $i$ and $j$ in cut-off radius via \cref{eq:eam_force_ij}.
    For each atom $i$, traverse all its neighbor atoms $j$, add force contributed by $j$ to atom $i$.
    Then we can obtain the total force for each atom.
\end{enumerate}

\subsection{Sunway TaihuLight and SW26010 Many-Core Processor}
Sunway TaihuLight \cite{fu_sunway_2016,yang_application_2018} is a supercomputer with a peak performance of 125.3 PFlops
and a sustained Linpack performance of 93 PFlops.
It was manufactured by \emph{National Research Center of Parallel Computer Engineering and Technology (NRCPC)}, and is hosted at National Supercomputing Center in Wuxi.
It comprises 40960 homegrown SW26010 many-core processors \cite{yang_application_2018}
which are connected by a 2-level fat-tree topology network. 

The SW26010 processor uses on-chip heterogeneous many-core architecture.
A SW26010 consists of four core groups (CGs) that are connected via the network on chip (NoC) and a system interface (SI) \cite{noauthor_homegrown_2015}.
A CG consists of a management processing element (MPE), a computing processing element (CPE) cluster, and a memory controller (MC).
Every CG has its own memory space, and the main memory is connected to the MPE and CPE cluster through the MC. 
Each CPE cluster contains 64 CPEs that are arranged in an $8 \times 8$ grid.
The SI is used to connect the processor itself and other devices outside.

The MPE and CPE are both 64-bit reduced instruction set computer (RISC) cores.
CPE is simpler than MPE and can only run in user mode.
Moreover, CPE is a core with only \SI{64}{KiB} user-controlled local device memory (LDM),
while MPE is connected with a 8 GB DDR3 memory via MC.
For memory accessing, CPE can access main memory directly by global load/store instructions (gld/gst),
or by direct memory access (DMA) which can be much faster than gls/gst.
At CPE cluster level, each CPE has 8 column communication buses and 8 row communication buses
to provide fast register communication across the $8 \times 8$ CPE grid,
which is a significant capability to share data inside CPE cluster.



\subsection{Related Work}
Many molecular dynamics programs were developed by numerous research teams.
Sandia National Labs developed LAMMPS \cite{plimpton1993fast} for classical molecular dynamics simulation.
However, the overhead of maintaining the neighbor list is very high during a large-scale simulation (e.g. $10^{10}$ particles).
Duan et al.\cite{duan_redesigning_2018} adapted LAMMPS for Sunway many-core architecture.
The new program achieved over 2.43 PFlops performance for a Tersoff simulation using 16,384 nodes.
However it only supports L-J potential and Tersoff potential, rather than the EAM potential that is more suitable for metallic systems.
GROMACS\cite{berendsen1995gromacs} is a widely-used molecular dynamic program in chemical and biomolecular.
SW\_GROMACS\cite{zhang_sw_gromacs_2019} accelerated GROMACS on Sunway TaihuLight supercomputer.
\new{Besides GROMACS, another significant MD software package for biomolecular simulation is NAMD \cite{phillips_namd_2002},
it is written using the Charm++ parallel programming model \cite{kale_charm_1993} and is the recipient of Gordon Bell award in 2002.}
ls1 Mardyn\cite{niethammer_ls1_2014} is another massively parallel molecular dynamics program
using L-J potential\cite{lennard1931cohesion} to solve the interaction among particles.
Its main target is the simulation of thermodynamics and nanofluidics. 
It achieved a $2.1 \times 10^{13}$-particle simulation on Hazel Hen supercomputer by using 172,032 CPU cores\cite{tchipev_twetris:_2019}.
However, both GROMACS and ls1 Mardyn cannot simulate the metallic systems with EAM potential.
Crystal MD\cite{hu_crystal_2017,hu_kernel_2017} is a parallel MD program for metal with BCC structure.
It carried out a four-trillion-atom simulation on Sunway TaihuLight supercomputer.
\new{SPaSM \cite{germann_25_nodate} is a MD implementations and can achieve simulation of more \change{than} 100 billions on BlueGeneL.}
But there is a lack of functionality of cascade collision simulation for both CrystalMD and SPaSM.

To support indexing of neighbor atoms within a cut-off radius, several classical data structures were proposed,
such as the neighbor list (or verlet lists)\cite{verlet_computer_1967} used in LAMMPS \new{, NAMD and GROMACS},
and link cell \cite{hockney_quiet_1974, niethammer_ls1_2014} used in ls1 Mardyn\new{ and SPaSM}.
In the neighbor list, each atom will create and maintain a reference to its neighbor atoms 
and store them in a list.
The neighbor list requires a lot of memory.
Link cell splits the simulation box into a number of cells.
In force computation step, for any atom $a$ in cell $c$,
link cell must traverse all atoms in cell $c$ and all atoms in neighbor cells of $c$.
which may index many unnecessary atoms out of cut-off radius.
Thus new design of data structure for efficient atoms storing and quick neighbor atoms indexing is essential.

\section{Hash Based Data Structure} \label{sec:data-structure}

\begin{figure}
    \centering
    \begin{adjustbox}{scale=0.76,center}
        \begin{tikzpicture}[->,>=stealth,level/.style={sibling distance = 2cm/#1,level distance = 1.0cm}]
  \tikzset{
    treenode/.style = {align=center, inner sep=0pt, text centered,
      font=\sffamily},
    arn_n/.style = {treenode, circle, white, font=\sffamily\bfseries, draw=black,
      fill=black, text width=1.5em},
    arn_r/.style = {treenode, circle, red, draw=red, 
      text width=1.5em, very thick},
    arn_x/.style = {treenode, rectangle, draw=black,
      minimum width=0.5em, minimum height=0.5em}
  }

    \pgfmathsetmacro{\X}{10.0}
    \pgfmathsetmacro{\Y}{8.0}
    \pgfmathsetmacro{\R}{0.17}
    \pgfmathsetmacro{\RNG}{0.22}


    \tikzstyle{compute} = [draw, shape=rectangle, fill=red!60, draw=red!60, thick, opacity=1,inner sep=0pt] 
    \tikzstyle{box} = [draw, shape=rectangle, fill=blue!30, draw=blue!60, thick, opacity=1,inner sep=0pt]
    \tikzstyle{box_text} = [midway,white]
    \tikzstyle{hash_array_arrow_text} = [midway, below, sloped, blue!80]
    \tikzstyle{hash_clash_arrow_text} = [hash_array_arrow_text,above, red!80]
    
    \tikzstyle{lat_node} = [fill=orange!60, draw=orange!60]
    \tikzstyle{empty_lat} = [fill=orange!60,fill opacity=0.1, draw=orange!60]
    \tikzstyle{clash_lat} = [fill=red!65,fill opacity=1, draw=red!60]

    \draw[very thin, color=lightgray] (0,0) grid (9.5, 7.5);

    \foreach \y in {0,...,3} {
        \foreach \x in {0,...,4} {
            \coordinate (l\x\y) at (2*\x+rand*\RNG, 2*\y+rand*\RNG);
            \coordinate (h\x\y) at (2*\x+1+rand*\RNG, 2*\y+1+rand*\RNG);
        };
    };

    \draw[lat_node] (l00) circle[radius=\R];
    \draw[empty_lat] (h00) circle[radius=\R];
    \draw[lat_node] (l10) circle[radius=\R];
    \draw[lat_node] (h10) circle[radius=\R];
    \draw[lat_node] (l20) circle[radius=\R];
    \draw[lat_node] (h20) circle[radius=\R];
    \draw[lat_node] (l30) circle[radius=\R];
    \draw[lat_node] (h30) circle[radius=\R];
    \draw[lat_node] (l40) circle[radius=\R];
    \draw[lat_node] (h40) circle[radius=\R];
    \draw[lat_node] (l01) circle[radius=\R];
    \draw[lat_node](h01) circle[radius=\R];
    \draw[lat_node] (l11) circle[radius=\R];
    \draw[lat_node] (h11) circle[radius=\R];
    \draw[lat_node] (l21) circle[radius=\R];
    \draw[lat_node] (h21) circle[radius=\R];
    \draw[empty_lat] (l31) circle[radius=\R];
    \draw[lat_node] (h31) circle[radius=\R];
    \draw[lat_node] (l41) circle[radius=\R];
    \draw[lat_node] (h41) circle[radius=\R];
    \draw[lat_node] (l02) circle[radius=\R];
    \draw[lat_node] (h02) circle[radius=\R];
    \draw[lat_node] (l12) circle[radius=\R];
    \draw[lat_node] (h12) circle[radius=\R];
    \draw[lat_node] (l22) circle[radius=\R];
    \draw[empty_lat] (h22) circle[radius=\R];
    \draw[lat_node] (l32) circle[radius=\R];
    \draw[lat_node] (h32) circle[radius=\R];
    \draw[lat_node] (l42) circle[radius=\R];
    \draw[lat_node] (h42) circle[radius=\R];
    \draw[lat_node] (l03) circle[radius=\R];
    \draw[lat_node] (h03) circle[radius=\R];
    \draw[lat_node] (l13) circle[radius=\R];
    \draw[lat_node] (h13) circle[radius=\R];
    \draw[lat_node] (l23) circle[radius=\R];
    \draw[lat_node] (h23) circle[radius=\R];
    \draw[lat_node] (l33) circle[radius=\R];
    \draw[lat_node] (h33) circle[radius=\R];
    \draw[lat_node] (l43) circle[radius=\R];
    \draw[lat_node] (h43) circle[radius=\R];

    \pgfmathsetmacro{\BH}{0.618} 
    \node [below] at (0.0+3.5,-3) {hash array}; 
    \draw[box] (0,-3) rectangle ++(1,\BH) node[box_text]{0};
    \draw[box] (0+1,-3) rectangle ++(1,\BH) node[box_text]{1};
    \draw[box] (0+2,-3) rectangle ++(1,\BH) node[box_text]{2};
    \draw[box] (0+3,-3) rectangle ++(1,\BH) node[box_text]{3};
    \draw[box] (0+4,-3) rectangle ++(2,\BH) node[box_text]{$\cdots$};
    \draw[box] (0+6,-3) rectangle ++(1,\BH) node[box_text]{$np$};

    \draw[blue!70,->] (l00) -- (0+1/2,-3+\BH) node[hash_array_arrow_text] {\footnotesize{$hash(x,y,z)$}};
    \draw[blue!70,->] (h00) -- (0+1+1/2,-3+\BH) node[hash_array_arrow_text] {\footnotesize{$hash(x,y,z)$}};
    \draw[blue!70,->] (l10) -- (0+2+1/2,-3+\BH) node[hash_array_arrow_text, above] {\footnotesize{$hash(x,y,z)$}};
    \draw[blue!70,->] (h10) -- (0+3+1/2,-3+\BH) node[hash_array_arrow_text] {\footnotesize{$hash(x,y,z)$}};
    \node[blue!70,thick] at (5, -2) {\large{$\cdots$}};

    \node(clash_atom1) at (4.22, 2.23) {};
    \node(clash_atom2) at (6.13, 4.30) {};
    \node(clash_atom3) at (8.23, 4.18) {};
    \draw[clash_lat] (clash_atom1) circle[radius=\R];
    \draw[clash_lat] (clash_atom2) circle[radius=\R];
    \draw[clash_lat] (clash_atom3) circle[radius=\R];

    \node [below] at (9,-3) {hash clash};
    \node (tree_root) [arn_r] at (9,-0.5) {} 
    child{ node (tree_left) [arn_n] {} 
        child{ node [arn_r] {}} 
        child{ node [arn_x] {}}
    }
    child{ node (tree_right) [arn_n] {} 
        child{ node [arn_r] {}} 
        child{ node [arn_r] {}}
    };
    
    \draw[red!70,->] (clash_atom1) -- (tree_left) node[hash_clash_arrow_text] {\footnotesize{$hash(x,y,z)$}};
    \draw[red!70,->] (clash_atom2) -- (tree_root) node[hash_clash_arrow_text,below] {\footnotesize{$hash(x,y,z)$}};
    \draw[red!70,->] (clash_atom3) -- (tree_right) node[hash_clash_arrow_text] {\footnotesize{$hash(x,y,z)$}};

\end{tikzpicture}
    \end{adjustbox}
    \caption{Atoms data structure and \change{indexing} method}
    \label{fig:md_data_structure}
\end{figure}
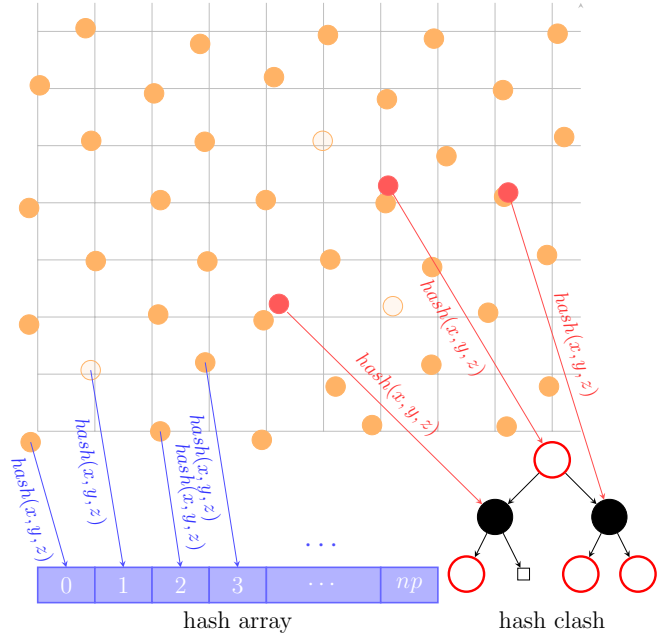

\begin{figure}
    \centering
    \begin{adjustbox}{scale=1.25,center}
        \begin{tikzpicture}[node distance=0cm]
    \tikzstyle{lattice} = [draw, shape=circle, fill=red!60, draw=none, inner sep=0pt, minimum size=3pt]
    
    \draw[help lines,step=0.5cm] (0,0) grid (4.75,4.75);

    \foreach \y in {0,...,4} 
    \draw [dotted] 
        (0,\y)         node [align=left, text=gray!80, below] { \tiny{$\left(0,\y \right)$} }
    --  (0.5,\y+0.5)   node [align=left, text=gray!80, below] { \tiny{$\left(1,\y \right)$} }
    --  (1,\y)         node [align=left, text=gray!60, below] { \tiny{$\left(2,\y \right)$} }
    --  (1.5,\y+0.5)   node [align=left, text=gray!60, below] { \tiny{$\left(3,\y \right)$} }
    --  (2,\y)         node [align=left, text=gray!60, below] { \tiny{$\left(4,\y \right)$} }
    --  (2.5,\y+0.5)   node [align=left, text=gray!60, below] { \tiny{$\left(5,\y \right)$} }
    --  (3,\y)         node [align=left, text=gray!60, below] { \tiny{$\left(6,\y \right)$} }
    --  (3.5,\y+0.5)   node [align=left, text=gray!60, below] { \tiny{$\left(7,\y \right)$} }
    --  (4,\y)         node [align=left, text=gray!60, below] { \tiny{$\left(8,\y \right)$} }
    --  (4.5,\y+0.5)   node [align=left, text=gray!60, below] { \tiny{$\left(9,\y \right)$} }
    --  (4.75,\y+0.25) node {};

    \foreach \y in {0,...,4} {
    \node [lattice] at (0,\y)  {};
    \node [lattice] at (0.5,\y+0.5)  {};
    \node [lattice] at (1,\y)  {};
    \node [lattice] at (1.5,\y+0.5)  {};
    \node [lattice] at (2,\y)  {};
    \node [lattice] at (2.5,\y+0.5)  {};
    \node [lattice] at (3,\y)  {};
    \node [lattice] at (3.5,\y+0.5)  {};
    \node [lattice] at (4,\y)  {};
    \node [lattice] at (4.5,\y+0.5)  {};
    };


\end{tikzpicture}
    \end{adjustbox}
    \caption{Lattice Offset Index for calculating neighbor atoms for BCC structure in 2d domain.
    The number in brackets under each lattice size is the coordinate of the corresponding lattice site}
    \label{fig:data_structure_offset}
\end{figure}
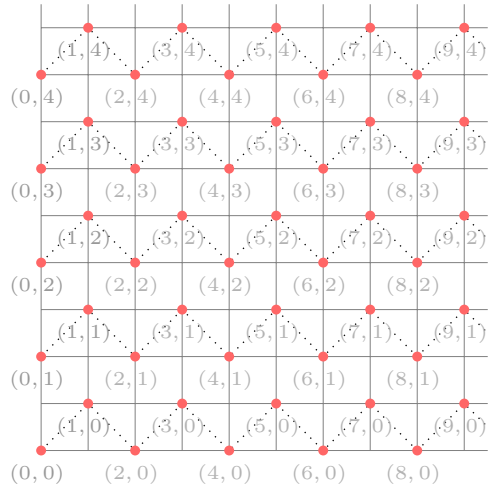

To support efficient atom storing and quick indexing of neighbor atoms,
a new hash-based data structure is designed for MISA-MD.
The hash-based data structure consists of two parts.
The first one is \MDDataStructure{} used to efficiently store atoms,
and the second one is \MDOffset{} targeting quick neighbor atom indexing.
In \MDDataStructureLowerCase{}, a lattice array will be established as a hash array, 
and a red-black tree will also be setup for storing elements with hash collision.
The hash function maps each atom to its nearest lattice in cartesian coordinate system.
\cref{fig:md_data_structure} illustrates this idea in a 2d simulation box.
For \MDOffsetLowerCase{}, the offset value of lattice id is pre-calculated for later neighbor atoms indexing.
To traverse neighbor atoms of a atom, the id of nearest lattice of the atom will be calculated first,
then we can obtain neighbor lattices by adding the nearest lattice id and offset lattice id.
\cref{fig:data_structure_offset} shows the lattice offset for calculating neighbor atoms.

The complete process of MD calculation based on BCC structure 
using \MDDataStructureLowerCase{} and \MDOffsetLowerCase{} data structure is list as following.
\paragraph{Hash Initialization}
At the beginning, a hash array for lattice will be allocated.
In our cascade collision simulation, body centered cubic (BCC) structure lattice is constructed for iron-based or tungsten-based material,
in which there is an additional lattice site located in the center of each cube unit.
Each element in hash array corresponds to a BCC lattice site.
For example, in the 2d simulation domain\footnote{In fact, it is a 3d domain in practice.},
the \textit{coordinate} of each lattice site can be recorded, as shown in \cref{fig:data_structure_offset},
and the coordinate can then be converted to hash array index by:
$$ index = 2 \times box_x \times y + x$$
where $box_x$ and $box_y$ are the lattice size of simulation box of current MPI process at $x$ and $y$ dimension, respectively.
and $x$ and $y$ are the corresponding lattice coordinates at each dimension.
In \cref{fig:data_structure_offset}, $box_x$ and $box_y$ are both equal to 5,
a lattice with coordinate $(3,4)$ is mapped to the 43rd ($2 \times 4 \times 5 + 3 =43$) element in hash array.
And in 3d simulation domain, the index can be calculated by:
\begin{equation}
    index = 2 \times box_x \times box_y \times z + 2 \times box_x \times y + x
\end{equation}
The advantage of this BCC lattice to index mapping method is that
we can store all lattice associated atoms in a compact memory mode.
We call the calculated index of hash array as \textit{lattice id}.
Meanwhile, the lattice id can also be converted to lattice coordinate easily.

The hash function can calculate \change{lattice} coordinate of the nearest lattice site of a atom and
return lattice id of the nearest lattice site.
Hash clash occurs when two distinct atom coordinates passed into the hash function produce identical lattice id outputs.
A empty red-black tree will be created for storing hash clash atoms and indexed by lattice id.
In this case, one of these atoms is stored in hash array and the left ones are stored in the red-black tree
when hash clash occurs.

Atoms are organized by this hash table.
In system initialization step of simulation, all atoms are placed at the corresponding lattice site.
In other words, there is no hash clash.
It is because that any two atoms must correspond to different lattice sites.
\change{Therefore}, all atoms data are stored in hash array at initialization step.

\paragraph{\MDOffset{} Initialization}
Under this regular lattice structure, list of offset lattice ids for indexing neighbor lattices in cut-off radius can be pre-calculated.

\paragraph{Iterate Neighbor Atoms}
We can calculate force or other interactions of each atoms in simulation domain
via \MDDataStructureLowerCase{} and \MDOffsetLowerCase{}.
\cref{alg:iterate_atoms_via_data_structure} show the algorithm of iterating all atoms and neighbor atoms
to calculate force.

\begin{algorithm}[!th]
    \caption{Iterate neighbor atoms to calculate force}
    \label{alg:iterate_atoms_via_data_structure}
    \KwInput{ $arr\_atoms$ - instance of hash array of \MDDataStructureLowerCase{}.}
    \KwInput{ $clash\_atoms$ - instance of hash clash of \MDDataStructureLowerCase{}.}
    \KwInput{ $\left(x_{start}, y_{start}, z_{start}\right), \left(x_{end}, y_{end}, z_{end}\right)$ - start and end lattice coordinate of the region for local atoms.}
    \KwInput{ $offset\_list$ - precalculated lattice id offset list}

    \For{$x \gets x_{start}, x_{end}$\rm ; $y \gets y_{start}, y_{end}$\rm; $z \gets z_{start}, z_{end}$}{
        $id \gets 2 \times box_x \times box_y \times z + 2 \times box_x \times y + x$ \;
        $atom \gets arr\_atoms[id]$ \;
        \If{$atom$ \rm is valid}{
            \For{\rm every $offset \in offset\_list$}{
                $atom\_nei \gets arr\_atoms[id + offset]$\;
                \If{$atom\_nei$ \rm is valid}{
                     $f \gets$ \Call{Force}{$atom$, $atom\_nei$} \;
                     $atom.f \gets atom.f + f$ \;
                }
            }
        }
    }

    \For{\rm every $atom \in clash\_atoms$}{
        $id \gets $ \Call{Hash}{$atom.x$, $atom.y$, $atom.z$} \;
        \If{$id$ \rm specified lattice is not in ghost area} {
            $atom\_near \gets arr\_atoms[id]$ \; 
            $f \gets$ \Call{Force}{$atom$, $atom\_near$} \;
            $atom.f \gets atom.f + f$ \;
            \For{\rm every $offset \in offset\_list$} {
                $atom\_nei \gets arr\_atoms[id + offset]$ \;
                \If{$atom\_nei$ \rm is valid}{
                    $f \gets$ \Call{Force}{$atom$, $atom\_nei$} \;
                    $atom.f \gets atom.f + f$ \;
                }
            }
        }
    }

    \For{\rm every $atom \in clash\_atoms$}{
        $id \gets $ \Call{Hash}{$atom.x$, $atom.y$, $atom.z$}  \;
        \If{$id$ \rm specified lattice is not in ghost area}{
            \For{\rm every $atoms\_cla \in clash\_atoms(id)$}{ 
                \If{$atom\_cla \neq atom$ }{
                    $f \gets$ \Call{Force}{$atom$, $atom\_cla$} \;
                    $atom.f \gets atom.f + f$ \;
                }
            }
            \For{\rm every $offset \in offset\_list$}{
                $id\_nei \gets id + offset$ \;
                \For{\rm every $atom\_cl \in clash\_atoms(id\_nei)$}{
                    $f \gets$ \Call{Force}{$atom$, $atom\_cl$} \;
                    $atom.f \gets atom.f + f$ \;
                }
            }
        }
    }
\end{algorithm}

Note that, for \change{obtaining} interaction contributions of neighbor atoms from neighbor MPI processes conveniently,
ghost region is involved.
The atoms in ghost region is also indexed by \MDDataStructureLowerCase{} together with local atoms located in real simulation domain.

In the algorithm, the calculation consists of three parts:
interactions of atoms from both hash array, interactions of atoms from hash array and hash clash,
and interactions of atoms from both hash clash.
In the first part of interactions computation, just iterate over the hash array and
traverse neighbor lattices of each lattice to obtain interaction contribution from neighbor atoms that are stored in hash array.
In interactions calculation of atoms from hash array and hash clash, 
iterate atoms in hash clash, if the atom is not in ghost area, find its nearest lattice via hash function,
then we can traverse neighbor atoms in hash array by the nearest lattice as center lattice.
At last, it falls into the interactions of atoms in hash clash.
It iterate the atoms that are not in ghost region.
For each atom $a$ in iteration, we can find the nearest lattice of this atom via hash function.
Then we can calculate ids of neighbor lattices by id of the nearest lattice and precalculated offset lattice ids,
and select atoms in hash clash with the neighbor lattice ids,
and add interactions contribution of these ones to atom $a$.


\paragraph{Hash Update}
After finishing each simulation step, the atoms coordinate can get changed,
thus updating of hash index is necessary.
\cref{alg:update_hash_index} presents the steps of hash updating.
The hash updating can be divided into two steps:
1) Traverse all elements (or atoms) in hash array.
For each valid element, re-calculate its lattice id via hash function.
If array index of the current element in hash array does not equal to the new calculated lattice id,
just copy this element to hash clash and tag this element as invalid.
2) Traverse all \change{of} atoms in hash clash.
For each atom in hash clash that is not located in ghost region, we can calculate its lattice id via hash function.
If the element in hash array indexed by the re-calculated lattice id is invalid,
then copy the atom data to hash array, set the element in hash array as valid, and remove this atom from hash clash. 

These two steps above in \cref{alg:update_hash_index} can make sure:
1) If two or more than two atoms have the identical hash value, 
only one of them is stored in hash array, and the left ones are stored to hash clash.
2) If there is a lattice id (index value in hash array) but no atom in simulation domain corresponds to this lattice id,
the element indexed by the lattice id in hash array must be invalid.
3) For each non-ghost atom in hash clash, if there is a invalid element in hash array that is indexed by the hash value of this atom,
the algorithm can remove the atom from hash clash and add it back to hash array.

\add{
The advantage of the hashed-based approach is that it can support quick neighbor searching 
and efficient atom storing, which use less memory compared than neighbor list method.
Denote $M$ as the number of neighbor in cut-off radius and $N$ as the number of atoms in simulation system.
The memory occupation can be divided as two parts: system atoms and neighbor indexing.
For system atoms, both the new hash-based approach and the neighbor list method 
need $O\left(N\right)$ memory to save system atoms.
But for neighbor indexing, the memory cost of the new hash-based approach will be $O\left(M\right)$, 
and memory cost of the neighbor list method will be $O(MN)$.
However, the disadvantage is also obvious.
The hash-based approach is specific to solids and would not work well for the system where the atoms are not on a regular lattice, such as liquids.
Because it will cost expensive computation to handle hash claims when the atoms are not on a regular lattice.
}
\begin{algorithm}[!th]
    \DontPrintSemicolon
    \caption{Hash updating}
    \label{alg:update_hash_index}
    \KwInput{ $arr\_atoms$ - instance of hash array of \MDDataStructureLowerCase{}.}
    \KwInput{ $clash\_atoms$ - instance of hash clash of \MDDataStructureLowerCase{}. }
    \KwInput{ $\left(x_{start}, y_{start}, z_{start}\right), \left(x_{end}, y_{end}, z_{end}\right)$ - start and end lattice coordinate of the region for local atoms.}

    \For{$x \gets x_{start}, x_{end}$}{
        \For{$y \gets y_{start}, y_{end}$}{
            \For{$z \gets z_{start}, z_{end}$}{
                 $id \gets 2 \times box_x \times box_y \times z + 2 \times box_x \times y + x$\;
                 $atom \gets arr\_atoms[id]$ \;
                \If{$atom$ \rm is invalid}{
                    continue \;
                }
                $id\_real \gets$ \Call{Hash}{$atom.x$,$atom.y$,$atom.z$}\;
                \If{$id \neq id\_real$} {
                    inset $atom$ into $clash\_atoms$ \;
                    set flag of $arr\_atoms[id]$ as invalid \;
                }
            }
        }
    }

    \For{\rm every $atom \in clash\_atoms$}{
        $id \gets $ \Call{Hash}{$atom.x$,$atom.y$,$atom.z$} \;
        \If{$id$ \rm specified lattice is not in ghost region}{
            \If{$arr\_atoms[id]$ \rm is invalid}{
                set flag of element $arr\_atoms[id]$ as valid \;
                $arr\_atoms[id] \gets atom$ \;
                remove $atom$ from $clash\_atoms$ \;
            }
        }
    }
\end{algorithm}

\section{Sunway Acceleration and Optimizations} \label{sec:sunway}
To achieve efficient MD simulation on large-scale clusters and make full use of the computation ability of SW26010 processor,
accelerating computation of MISA-MD is essential.
Results from profile tools show that the EAM potential computation in MISA-MD,
consume more than 80\% computation time.
In SW26010 processor, 256 CPEs in 4 core group contribute more than 90\% peak performance of the overall processor\cite{lin_evaluating_2018}.
Therefore, the code for accelerating EAM interaction computation on sunway CPEs is implemented.
In current version of MISA-MD, we only consider the acceleration for interaction of atoms from hash array
due to the fact of low rate of hash clash.

Moreover, it is of great significance to accelerate EAM computation on SW26010.
Because it is another significant approach to extend MD simulation to longer temporal scale besides quick neighbor atoms indexing.

\subsection{Potential Tables Storage and Interpolation}
In EAM computation, the $n \times \left(n + 1\right)/2 + n + n + n$ functions for $n$ different atom types
are determined by spline interpolation
from a table with dispersed values called ``potential table''.

For each to be determined function as well as its partial derivative,
$N$ equidistant dispersed points are presented in potential table.
For instance, to determine electronic density function of atom type $\beta$,
point set $P_\beta=\{(r_1,\rho_1), (r_2,\rho_2), (r_3,\rho_3), \cdots, (r_N,\rho_N) \}$ is given in table,
where $r_i$ is distance and $\rho_i$ is value of electronic density under distance $r_i$, $i=1,2,\cdots,N$.
We are expected to find a electronic density function $\rho_{\beta}(r)$ and
its partial derivative function $\dfrac{\partial \rho_{\beta} (r)}{\partial r}$
for $ r \in \left[r_1,r_N \right]$.

To perform interpolation on a point set $P=\{(x_1,y_1), (x_2,y_2), (x_3,y_3), \cdots, (x_n,y_N) \}$
with condition of $x_1 < x_2 < \cdots < x_N$ and $x_2 - x_1 = x_3 - x_2 = \cdots = x_N - x_{N-1} = h$,
a cubic polynomial function $S_i(x) = a_i(x-x_i)^3+b_i(x-x_i)^2+c_i(x-x_i)+d_i$
is selected for each segment specified by two adjacent point to approximate the undetermined function in spline interpolation.
Then we can calculate the coefficients $a_i$, $b_i$, $c_i$, and $d_i$ of $S_i(x)$ on each segment,
as well as coefficients $e_i$, $f_i$ and $g_i$ for partial derivative of $S_i(x)$:
$S_i'(x)=e_i(x-x_i)^2+f_i(x-x_i)+g_i$.
After each functions is determined, then these seven coefficients on each segment can be stored for later EAM computation.

In our cases, to simulate cascade collision of a pure Fe system using MISA-MD,
a potential table file from \cite{bonny_ternary_2009} is provided, which contains 3 potential tables
(one pair-wise potential table for Fe-Fe, one for embedding energy for Fe and one for electronic density for Fe).
Each potential table above contains \SI{5000}{} dispersed data values used for interpolation to
determine corresponding potential function and it partial derivative function.
This means that, we need at least \SI{820}{KiB} memory to store the interpolation coefficients for single-atom-type system
(3 potential tables $\times$ 7 coefficients $\times$ 5000 segments $\times$ 8 bytes per double precision floating point value),
which is much large than \SI{64}{KiB} LDM of CPE.
Further, if the target system is an alloy system containing multiple types of atoms, the storage of coefficients can be much larger.
Thus, the interpolation coefficients can not be loaded into the LDM of CPE at one time.

To overcome the small LDM of CPE, we copy the origin values of one table to LDM of CPE 
before using it, whose size is only \SI{39}{KiB}.
This thought is \change{inspired} by \cite{li_massively_2018}.
Then, all the coefficients of $S_i(x)$ and $S_i'(x)$ can be calculated on the fly
using the origin data and interpolation method.
But the shortcoming of this method is obvious.
It can only calculate single-atom-type system, 
because for alloy system, the LDM of CPE cannot hold all necessary origin data.

To deeply optimize the memory storage of origin potential table data,
we exchange order of offset list iteration and atoms iteration in \cref{alg:iterate_atoms_via_data_structure}
(move the offset list iteration to the outermost loop and move atoms iteration of x,y,z dimension to inner loop).
Because the origin potential table is associated with atoms distance $r$,
in each offset list iteration, we can obtain the minimum and maximum distance of neighbor atoms in atoms iteration \change{firstly},
and then we can only load part of the origin potential data that is in this distance range, rather than the whole origin data.
With partial potential data loading in each offset list iteration,
we can free up more memory space that can be used for atoms storage.
But the number of potential data copies from main memory to LDM can increase due to the partial potential data loading.

To overcome the large overlap of potential data copies, 
we divide the origin data of potential tables into some blocks 
and each CPE pre-load a block of origin potential tables at initialization step.
Then the origin potential tables are stored on CPEs distributedly.
And in latter offset list iteration,
the partial potential data is loaded from some specific CPEs via fast register communication, instead of main memory.

\subsection{Tasks Assignment on CPEs}
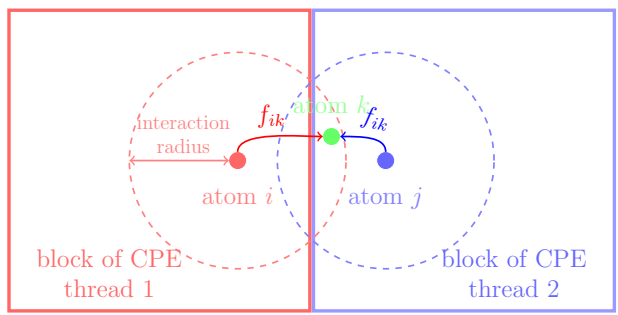
\begin{figure}
    \centering
    \begin{adjustbox}{scale=0.8,center}
        \begin{tikzpicture}[node distance=0cm]
    \pgfmathsetmacro{\B}{5.0}
    \pgfmathsetmacro{\BC}{\B+0.06}
    \pgfmathsetmacro{\R}{1.8}

    \tikzstyle{compute} = [draw, shape=rectangle, fill=red!60, draw=red!60, thick, opacity=1,inner sep=0pt] 
    \tikzstyle{dma} = [draw, shape=rectangle, fill=blue!30, draw=blue!60, thick, opacity=1,inner sep=0pt]
    \tikzstyle{compute_text} = [align=center,text width=3cm, minimum size=3pt, inner sep=0pt]

    \draw[ultra thick,red!60] (0, 0) -- ++(0,\B) -- ++(\B,0) -- ++(0,-\B) -- cycle;
    \draw[ultra thick,blue!40] (\BC, 0) -- ++(0,\B) -- ++(\B,0) -- ++(0,-\B) -- cycle;

    \node (lc)  at (\B-\R/1.5,\B/2) {};
    \node [below=0.15 of lc,red!50] {atom $i$};
    \draw[fill=red!60, draw=white] (lc) circle [radius=0.15];
    \draw[red!50,dashed, thick] (lc) circle [radius=\R];
    \draw[<->, thick, red!50] (lc) -- ++(-\R/2,0)node[above, align=center, text width=3cm, scale=0.8]{interaction \\ radius} -- ++(-\R/2,0); 

    \node (rc)  at (\BC+\R/1.5,\B/2) {};
    \node [below=0.15 of rc,blue!50] {atom $j$};
    \draw[fill=blue!60, draw=white] (rc) circle [radius=0.15];
    \draw[blue!50, dashed, thick] (rc) circle [radius=\R];

    \node (atom_k) at (\BC+0.3,\B/2+0.4) {};
    \draw[fill=green!60, draw=white] (atom_k) circle [radius=0.15];
    \node [above=0.1 of atom_k, green!40] {atom $k$};

    \draw[->,thick, red] (lc) .. controls +(up:0.5) and +(left:0.5) .. node[above,sloped] {$f_{ik}$} (atom_k);
    \draw[->,thick, blue] (rc) .. controls +(up:0.4) and +(right:0.5) .. node[above,sloped] {$f_{ik}$} (atom_k);

    \node[compute_text,red!60] at (\B/3,\B/8)  {block of CPE thread 1};
    \node[compute_text,blue!60] at (\BC+\B-\B/3,\B/8)  {block of CPE thread 2};

\end{tikzpicture}
    \end{adjustbox}
    \caption{Write Conflict of \change{CPE threads}}
    \label{fig:write_conflict_of_CPE_thread}
\end{figure}

In fact, Newton's third law can be applied for force calculation.
As described at the end of \cref{sec:eam_computation_steps}, thanks to Newton's third law,
the amount of calculation of atoms' electronic density contribution at step 1) and
and force contribution at step 5) as well as pair-wise potential at step 3) and derivative of electronic density at step 4)
can halve.
But the problem of threads' write conflict on sunway CPEs must be solved before involving this trick.
When we assign computation tasks of one MPI processor to 64 CPE \change{threads}
\add{(In sunway native programming, one CPE can launch a light-weight working thread using Athread library \cite{cai_openacc_2018}.)},
spatial decomposition of simulation domain is achieved.
We may split the domain owned by the MPI processor into 64 blocks,
and each CPE \change{thread} would carry out the EAM calculation of one block of them.
But the write conflict will occur when writing calculated results back to main memory on MPE
if there is common neighbor atom for two adjacent CPE \change{threads},
as shown in \cref{fig:write_conflict_of_CPE_thread}.

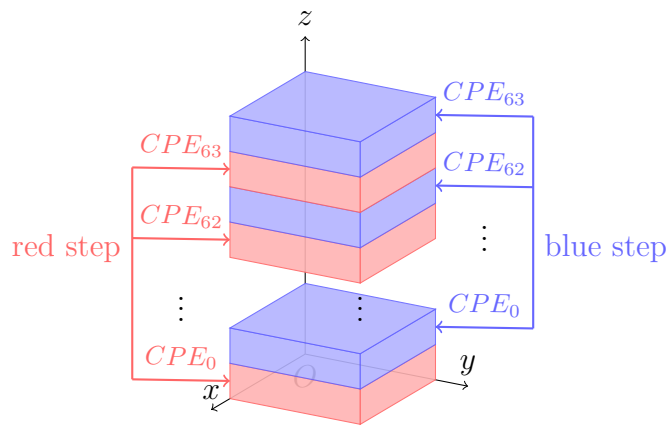
\begin{figure}
    \centering
    \begin{adjustbox}{scale=1,center}
        \tdplotsetmaincoords{70}{120}
\pgfmathsetmacro{\Width}{2}
\pgfmathsetmacro{\Height}{2}
\pgfmathsetmacro{\Depth}{3}

\pgfmathsetmacro{\D}{0.5}
\pgfmathsetmacro{\B}{1.5}

\pgfmathsetmacro{\X}{2.5}
\pgfmathsetmacro{\Y}{2.5}
\pgfmathsetmacro{\Z}{4.5}

\begin{tikzpicture}[tdplot_main_coords]
    \coordinate (O) at (0,0,0);
    \coordinate (A) at (0,\Height,0);
    \coordinate (B) at (0,\Height,\Depth);
    \coordinate (C) at (0,0,\Depth);
    \coordinate (D) at (\Width,0,0);
    \coordinate (E) at (\Width,\Height,0);
    \coordinate (F) at (\Width,\Height,\Depth);
    \coordinate (G) at (\Width,0,\Depth);

\draw [black,->] (0,0,0)node[anchor=north]{$O$} -- (\X,0,0) node[anchor=south]{$x$};
\draw [black,->] (0,0,0) -- (0,\Y,0) node[anchor=south]{$y$};
\draw [black,->] (0,0,0) -- (0,0,\Z) node[anchor=south]{$z$};

\draw[red!60,fill=red!30,opacity=0.9] (\Width,0,0) -- ++(0,\Height,0) -- ++(0,0,\D) -- ++(0,-\Height,0) -- cycle;
\draw[red!60,fill=red!30,opacity=0.9] (0,\Height,0) -- ++(\Width,0,0) -- ++(0,0,\D) -- ++(-\Width,0,0) -- cycle;

\draw[blue!60,fill=blue!30,opacity=0.9] (\Width,0,\D) -- ++(0,\Height,0) -- ++(0,0,\D) -- ++(0,-\Height,0) -- cycle;
\draw[blue!60,fill=blue!30,opacity=0.9] (0,\Height,\D) -- ++(\Width,0,0) -- ++(0,0,\D) -- ++(-\Width,0,0) -- cycle;
\draw[blue!60,fill=blue!30,opacity=0.9] (0,0,2*\D) -- ++(\Width,0,0) -- ++(0,\Height,0) -- ++(-\Width,0,0) -- cycle;

\node [black] at (\Width,\Height,3.5*\D)  {$\vdots$};


\draw[red!60,fill=red!30,opacity=0.9] (\Width,0,4*\D) -- ++(0,\Height,0) -- ++(0,0,\D) -- ++(0,-\Height,0) -- cycle;
\draw[red!60,fill=red!30,opacity=0.9] (0,\Height,4*\D) -- ++(\Width,0,0) -- ++(0,0,\D) -- ++(-\Width,0,0) -- cycle;

\draw[blue!60,fill=blue!30,opacity=0.9] (\Width,0,5*\D) -- ++(0,\Height,0) -- ++(0,0,\D) -- ++(0,-\Height,0) -- cycle;
\draw[blue!60,fill=blue!30,opacity=0.9] (0,\Height,5*\D) -- ++(\Width,0,0) -- ++(0,0,\D) -- ++(-\Width,0,0) -- cycle;

\draw[red!60,fill=red!30,opacity=0.9] (\Width,0,6*\D) -- ++(0,\Height,0) -- ++(0,0,\D) -- ++(0,-\Height,0) -- cycle;
\draw[red!60,fill=red!30,opacity=0.9] (0,\Height,6*\D) -- ++(\Width,0,0) -- ++(0,0,\D) -- ++(-\Width,0,0) -- cycle;

\draw[blue!60,fill=blue!30,opacity=0.9] (\Width,0,7*\D) -- ++(0,\Height,0) -- ++(0,0,\D) -- ++(0,-\Height,0) -- cycle;
\draw[blue!60,fill=blue!30,opacity=0.9] (0,\Height,7*\D) -- ++(\Width,0,0) -- ++(0,0,\D) -- ++(-\Width,0,0) -- cycle;
\draw[blue!60,fill=blue!30,opacity=0.9] (0,0,8*\D) -- ++(\Width,0,0) -- ++(0,\Height,0) -- ++(-\Width,0,0) -- cycle;

\draw[red!60,thick,->] (\Width,-\B,0*\D) --  (\Width,0,0.5*\D) node[above,midway,sloped] {\footnotesize{$CPE_0$}};
\draw[red!60,thick,->] (\Width,-\B,4*\D) --  (\Width,0,4.5*\D) node[above,midway,sloped] {\footnotesize{$CPE_{62}$}};
\draw[red!60,thick,->] (\Width,-\B,6*\D) --  (\Width,0,6.5*\D) node[above,midway,sloped] {\footnotesize{$CPE_{63}$}};
\draw[red!60,thick,-] (\Width,-\B,0*\D) -- ++(0,0,6*\D)node[above left,midway]{red step};
\node [black] at (\Width,-\B/2,2.5*\D)  {$\vdots$};

\draw[blue!60,thick,->] (0,\Height+\B,2*\D) --  (0,\Height,1.5*\D) node[above,midway,sloped] {\footnotesize{$CPE_0$}};
\draw[blue!60,thick,->] (0,\Height+\B,6*\D) --  (0,\Height,5.5*\D) node[above,midway,sloped] {\footnotesize{$CPE_{62}$}};
\draw[blue!60,thick,->] (0,\Height+\B,8*\D) --  (0,\Height,7.5*\D) node[above,midway,sloped] {\footnotesize{$CPE_{63}$}};
\draw[blue!60,thick,-] (0,\Height+\B,2*\D) -- ++(0,0,6*\D)node[below right,midway]{blue step};

\node [black] at (0,\Height+\B/2,4.5*\D)  {$\vdots$};

\end{tikzpicture}
    \end{adjustbox}
    \caption{Coloring \change{CPE threads}}
    \label{fig:coloring_of_CPE_thread}
\end{figure}

To overcome this difficulty, we proposed a coloring method for sunway CPE \change{threads},
which is \change{inspired} by SDC (Spatial Decomposition Colouring) method\cite{liu_efficient_2011,hu_efficient_2009}  
used in OpenMP program model on multi-core platforms.
\cref{fig:coloring_of_CPE_thread} shows an example of the coloring result.
In coloring method, we split the domain space into 128 blocks in z dimension, 
where 128 is twice of the number of CPEs,
and each block is colored as red or blue and make sure there is no adjacent blocks with the same color.
Then the calculation is divided into two steps to avoid write conflict.
In the first step, each CPE \change{thread} is assigned a block with red color,
and the next step, each CPE \change{thread} is assigned a block with blue color.
With these two steps, all blocks can be calculated without write conflict.

\subsection{Double-Buffer DMA and Data Reuse Strategies}
\begin{figure}
    \centering
    \begin{adjustbox}{scale=0.6,center}
        \begin{tikzpicture}[node distance=0cm]
    \pgfmathsetmacro{\H}{1.0}
    \pgfmathsetmacro{\S}{\H+0.2}
    \pgfmathsetmacro{\W}{1.75}
    \pgfmathsetmacro{\WD}{1.0}
    \pgfmathsetmacro{\Offset}{2.0}
    \pgfmathsetmacro{\OffDot}{1.5}

    \tikzstyle{compute} = [draw, shape=rectangle, fill=red!60, draw=red!60, thick, opacity=1,inner sep=0pt] 
    \tikzstyle{dma} = [draw, shape=rectangle, fill=blue!30, draw=blue!60, thick, opacity=1,inner sep=0pt]
    \tikzstyle{compute_text} = [align=center,text width=1.6cm, minimum size=3pt, inner sep=0pt, white!60]

    \node at (0.2,\H/2)  {Buffer$_2$};
    \node at (0.2,\S+\H/2)  {Buffer$_1$};
 
    \draw[fill=black,thick,->] (1, 2.5) -- ++(6.6,0)node[fill=white]{time} -- ++(6.6,0);

    \draw[compute] (1+\WD,\S) -- ++(\W,0) -- ++(0,\H) -- ++(-\W,0) -- cycle;
    \draw[compute] (1+\WD+\W,0) -- ++(\W,0) -- ++(0,\H) -- ++(-\W,0) -- cycle;
    \draw[compute] (1+\WD+2*\W,\S) -- ++(\W,0) -- ++(0,\H) -- ++(-\W,0) -- cycle;
    \draw[compute] (1+\WD+3*\W,0) -- ++(\W,0) -- ++(0,\H) -- ++(-\W,0) -- cycle;

    \draw[compute] (\Offset+1+\WD+4*\W,\S) -- ++(\W,0) -- ++(0,\H) -- ++(-\W,0) -- cycle;
    \draw[compute] (\Offset+1+\WD+5*\W,0) -- ++(\W,0) -- ++(0,\H) -- ++(-\W,0) -- cycle;

    \draw[dma] (1+0,\S) -- ++(\WD,0) -- ++(0,\H) -- ++(-\WD,0) -- cycle;
    \draw[dma] (1+0,0) -- ++(\WD,0) -- ++(0,\H) -- ++(-\WD,0) -- cycle;
    \draw[dma] (1+\WD+\W,\S) -- ++(\WD,0) -- ++(0,\H) -- ++(-\WD,0) -- cycle;
    \draw[dma] (1+\WD+2*\W,0) -- ++(\WD,0) -- ++(0,\H) -- ++(-\WD,0) -- cycle;
    \draw[dma] (1+\WD+3*\W,\S) -- ++(\WD,0) -- ++(0,\H) -- ++(-\WD,0) -- cycle;
    \draw[dma] (1+\WD+4*\W,0) -- ++(\WD,0) -- ++(0,\H) -- ++(-\WD,0) -- cycle;

    \draw[dma] (\Offset+1+\WD+3*\W, \S) -- ++(\WD,0) -- ++(0,\H) -- ++(-\WD,0) -- cycle;
    \draw[dma] (\Offset+1+\WD+4*\W, 0) -- ++(\WD,0) -- ++(0,\H) -- ++(-\WD,0) -- cycle;

    \draw (\OffDot+1+\WD+3*\W, \S+\H/2) node[black]{$\cdots$};
    \draw (\OffDot+1+\WD+4*\W, 0+\H/2) node[black]{$\cdots$};
 
    \draw  (1+\WD+\W/2, \S+\H/2) node[compute_text]{EAM compute};
    \draw  (1+\WD+\W+\W/2, 0+\H/2) node[compute_text]{EAM compute};
    \draw  (1+\WD+2*\W+\W/2, \S+\H/2) node[compute_text]{EAM compute};
    \draw  (1+\WD+3*\W+\W/2, 0+\H/2) node[compute_text]{EAM compute};
    \draw (\Offset+1+\WD+4*\W+\W/2, \S+\H/2) node[compute_text]{EAM compute};
    \draw (\Offset+1+\WD+5*\W+\W/2, 0+\H/2) node[compute_text]{EAM compute};

    \draw (1+0+\WD/2, \S+\H/2) node[black!60] {DMA};
    \draw (1+0+\WD/2, 0+\H/2) node[black!60] {DMA};
    \draw (1+\WD+\W+\WD/2, \S+\H/2) node[black!60] {DMA};
    \draw (1+\WD+2*\W+\WD/2, 0+\H/2) node[black!60] {DMA};
    \draw (1+\WD+3*\W+\WD/2, \S+\H/2) node[black!60] {DMA};
    \draw (1+\WD+4*\W+\WD/2, 0+\H/2) node[black!60] {DMA};
    \draw (\Offset+1+\WD+3*\W+\WD/2, \S+\H/2) node[black!60] {DMA};
    \draw (\Offset+1+\WD+4*\W+\WD/2, 0+\H/2) node[black!60] {DMA};

\end{tikzpicture}
    \end{adjustbox}
    \caption{Double-buffer DMA strategy}
    \label{fig:sunway_double_buffer}
\end{figure}
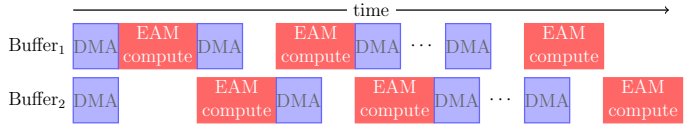

Naturally, the domain block assigned to a CPE may contain a large number of atoms which
can not be \change{transmitted} to CPE at one time due to the restriction of \SI{64}{KiB} LDM.
Thus, the domain block is \change{split} into numbers of small data block for CPE computation.
In traditional method, CPE would request a memory block of atoms data from main memory,
then it start computation once the atoms data transmission finish.
After computation completes, atoms data will put back to main memory and then
prepare to access next data block for next round computation repeatedly
until all data blocks assigned to this CPE is completed.
We found that the repeatedly DMA data accessing has a high overhead, which may leading a poor performance.
Therefore, a double-buffer DMA strategy is taken to overlap the DMA data transmission and EAM potential computation,
as shown in \cref{fig:sunway_double_buffer}.
In double-buffer DMA strategy, the LDM is divided into three blocks,
one is for origin potential data, the second and the third ones are both buffer memory for each other.
Expect for the initial DMA getting and the last DMA putting operation,
when performing computation of current round, the DMA putting operation of previous round or
the DMA getting operation of next round can be performed simultaneously.
By overlaping DMA data transmission and EAM potential computation,
the computation efficiency on CPEs is greatly improved.

Double buffer strategy can overlap DMA and computation effectively.
To reduce the overhead of DMA further, another optimization, called data reuse, is applied.
Considering atoms data block of current calculation round,
a part of them can be actually the ghost atoms data of next calculation round,
and its ghost data can also be the `real' atom data of next round.
Thus, we can keep the these two part of data in the LDM of CPE for usage of next round calculation,
which can reduces the amount of data transferred to LDM in each DMA operation.

\section{Evaluation} \label{sec:results}
This section presents the evaluation on single core group performance, scalability, and accuracy of cascade collision simulations.
As a comparison, we also compared memory usage and performance with LAMMPS on Sunway and Intel platforms.

\subsection{Accuracy and Case Study}

\begin{figure}
    \centering
    \begin{subfigure}[]{0.49\textwidth}
        \begin{adjustbox}{scale=1,center}
            \includegraphics[width=\textwidth]{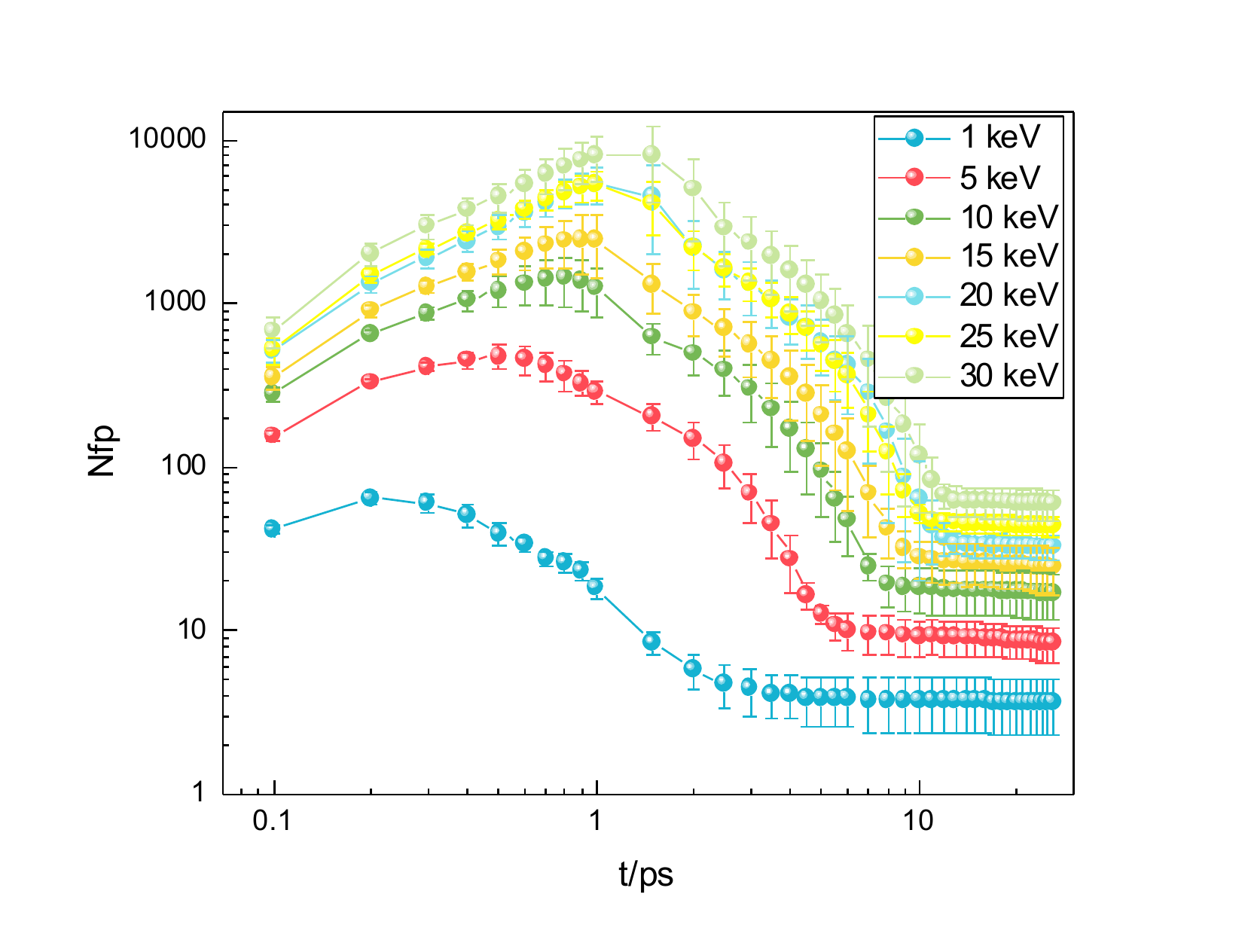}
        \end{adjustbox}
        \caption{Evolution of number of displacement cascades' defects for different PKA energy from MISA-MD }
        \label{fig:frenkel_defects_misa-md}
    \end{subfigure}
    \begin{subfigure}[]{0.49\textwidth}
        \begin{adjustbox}{scale=1,center}
            \includegraphics[width=\textwidth]{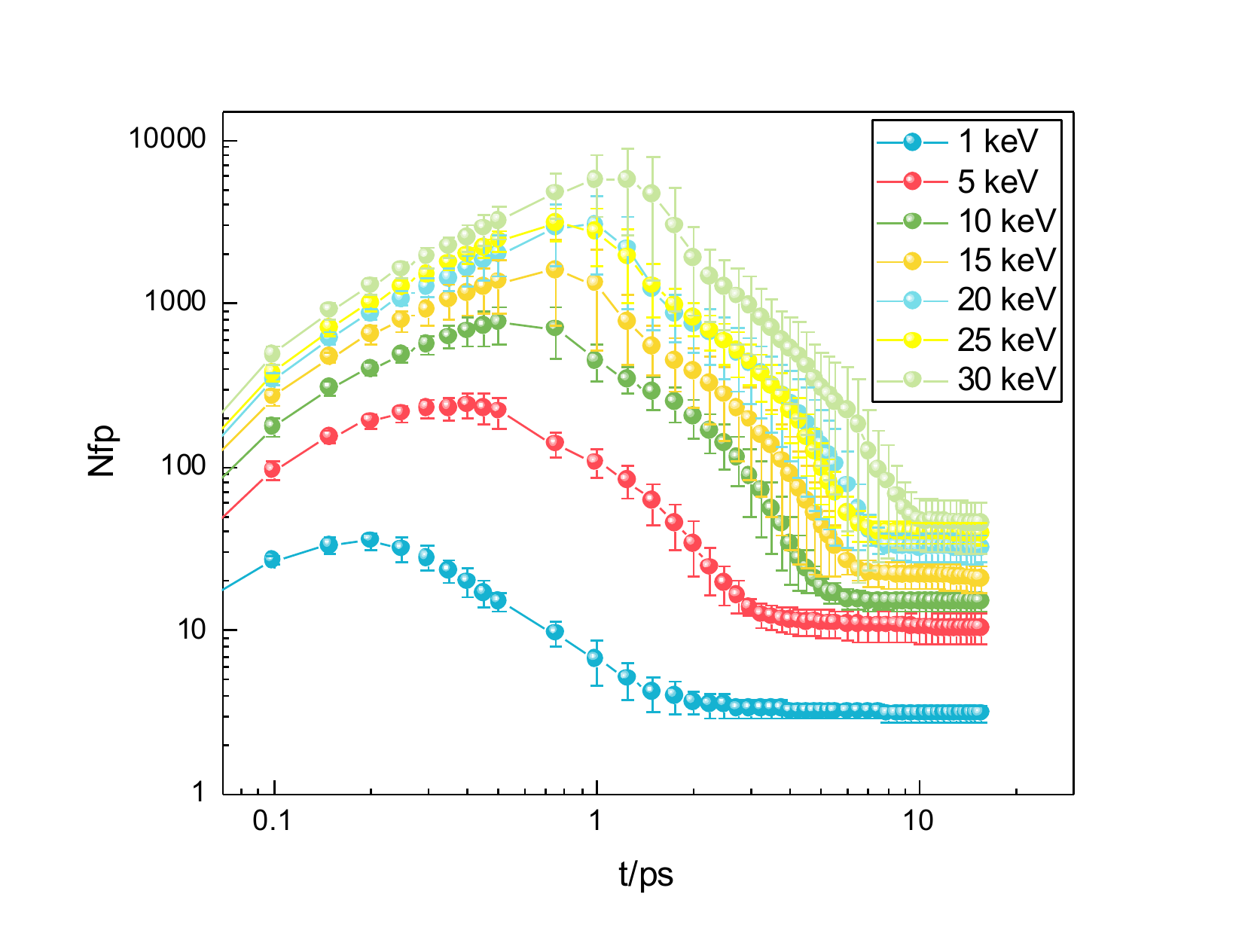}
        \end{adjustbox}
        \caption{Evolution of number of displacement cascades' defects for different PKA energy from LAMMPS}
        \label{fig:frenkel_defects_lmp}
    \end{subfigure}
    \caption{Evolution of number of displacement cascades' defects for different PKA energy at 600K temperatures}
    \label{fig:frenkel_number_evolution}
\end{figure}

In order to validate the accuracy of MISA-MD,
we have performed several cascade collision simulations of different PKA energies and directions.
In our cases, the PKA energy settings are \SI{1.0}{\kilo\electronvolt}, \SI{5.0}{\kilo\electronvolt}, \SI{10.0}{\kilo\electronvolt},
\SI{15.0}{\kilo\electronvolt}, \SI{20.0}{\kilo\electronvolt}, \SI{25.0}{\kilo\electronvolt} and \SI{30.0}{\kilo\electronvolt},
and PKA directions includes $\langle 1,2,2 \rangle$, $\langle 1,3,5 \rangle$ and $\langle 2,3,5 \rangle$.
In each simulation instance, a $80 a_0 \times 80 a_0 \times 80 a_0$ (where $a_0$ is lattice constant) simulation box containing \SI{1024000}{} Fe atoms is involved,
and the primary knock-on atom is placed at position of $(40a_0,40a_0,40a_0)$ (\change{center} of simulation box) with a corresponding velocity associated with PKA energy.
For each PKA energy and each PKA, simulations are performed three times to obtain the average results and
analyze the number of frenkel pairs in simulating box over evolution time.
As shown in \cref{fig:frenkel_defects_misa-md}.
Besides, we also tested on LAMMPS using EAM potential with the same input parameters.
\cref{fig:frenkel_defects_lmp} shows the evolution of frenkel pairs numbers for different PKA energies at 600K temperatures from simulation results of LAMMPS.

\begin{figure}
    \centering
    \begin{subfigure}[]{0.24\textwidth}
        \begin{adjustbox}{scale=0.8,center}
            \includegraphics[width=\textwidth]{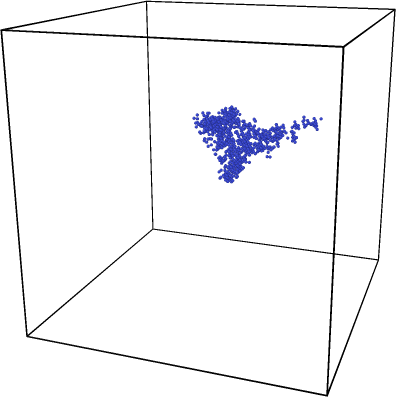}
        \end{adjustbox}
        \caption{system configuration after \SI{0.1}{\ps} of PKA generation within 424 Frenkel pairs.}
        \label{fig:frenkel_defects_1}
    \end{subfigure}
    \begin{subfigure}[]{0.24\textwidth}
        \begin{adjustbox}{scale=0.8,center}
            \includegraphics[width=\textwidth]{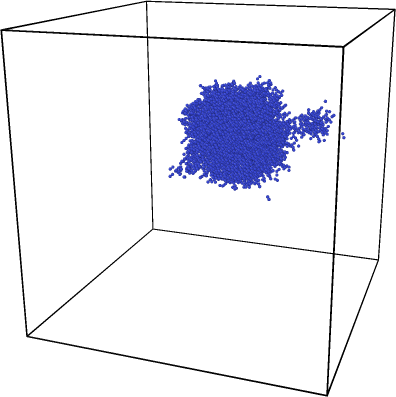}
        \end{adjustbox}
        \caption{system configuration after \SI{0.8}{\ps} of PKA generation within 3307 Frenkel \change{pairs}.}
        \label{fig:frenkel_defects_2}
    \end{subfigure}
    \begin{subfigure}[]{0.24\textwidth}
        \begin{adjustbox}{scale=0.8,center}
            \includegraphics[width=\textwidth]{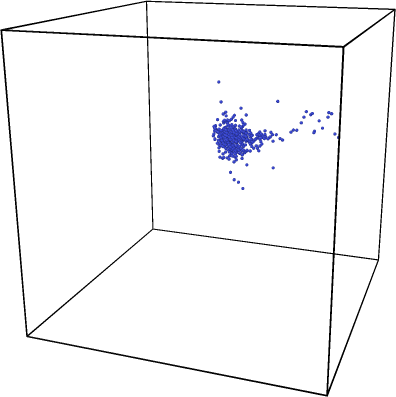}
        \end{adjustbox}
        \caption{system configuration after \SI{5.5}{\ps} of PKA generation within 190 Frenkel \change{pairs}.}
        \label{fig:frenkel_defects_3}
    \end{subfigure}
    \begin{subfigure}[]{0.24\textwidth}
        \begin{adjustbox}{scale=0.8,center}
            \includegraphics[width=\textwidth]{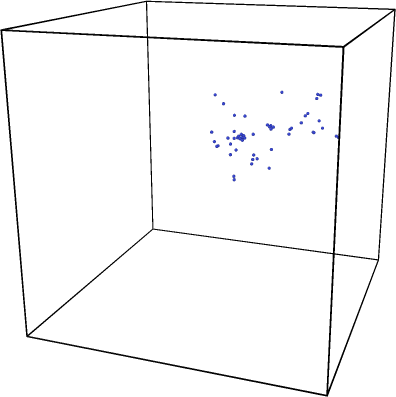}
        \end{adjustbox}
        \caption{system configuration after \SI{26.0}{\ps} of PKA generation within 26 Frenkel \change{pairs}.}
        \label{fig:frenkel_defects_4}
    \end{subfigure}
    \caption{MISA-MD visualization of system evolution at different time under PKA energy of 15kev and direction of $\langle 1,3,5 \rangle$.}
    \label{fig:frenkel_defects_visualization}
\end{figure}

Comparing \cref{fig:frenkel_defects_misa-md} and \cref{fig:frenkel_defects_lmp},
their evolution curves of frenkel pairs number analyzed from simulation results have the same trend under varies of PKA energies.
The number of frenkel pairs of MISA-MD and LAMMPS both grow to a peak rapidly after a short time of PKA generation
and then quench from peak value to stable value.
As a case of this process, \cref{fig:frenkel_defects_visualization} visualizes
process of cascade collision simulated by MISA-MD,
in which the spatial distributions of frenkel pairs at different evolution stages
under PKA energy of 15kev and incidence direction of $\langle 1,3,5 \rangle$ are presented.
Moreover, with a large PKA energy, the peak would appear later with a larger peak,
and a higher value of frenkel pairs survives at stable stage.
By comparing peak time, peak value, stable time and stable value of MISA-MD and LAMMPS quantitatively,
they are all \change{similar} and within acceptable tolerances. 
And the number of survived frenkel pairs pairs also meet the classical NRT model established by Norget \cite{norgett_proposed_1975}.

\subsection{Single Core Group Evaluation}
To test the computational efficiency of code accelerated by $8 \times 8$ CPEs,
we performed performance test on single core group with and without CPEs acceleration.
In this test case, a $200 a_0 \times 200 a_0 \times 256 a_0$ (where $a_0$ is lattice constant) simulation box was involved
and simulations of 10 time steps was performed.
\cref{tab:single_node_evaluation} shows the execution time and speedup of case \change{with} MPE only
and case with MPE plus 64 CPEs, 
which indicates our MISA-MD code obtains approximate 56 acceleration speedup with \change{computation} on CPEs.

\begin{table}[h]
    \caption{Speedup of MISA-MD on single core group with and without CPEs.}
    \begin{tabular}{lll}
        \hline
                 & MPE only & MPE+64CPEs  \\
        \hline
        Execution time (seconds)  & 1192.33  &    21.2939   \\
        Speedup    &  1 & 55.99 \\
        \hline
    \end{tabular}
    \label{tab:single_node_evaluation}
\end{table}

\subsection{Comparison with LAMMPS}
In this subsection, we compare the memory usage and computational performance of MISA-MD and LAMMPS
on Sunway and Intel platforms.
For LAMMPS, the current latest version, \textit{stable\_3Mar2020}, is selected to be compared.
As the EAM computation of LAMMPS does not supported Sunway CPEs architecture,
thus, we \change{perform} performance tests of LAMMPS and MISA-MD using MPEs only on Sunway platform.

\subsubsection{Memory usage}
The memory usage of MISA-MD and LAMMPS is compared at runtime to demonstrate the superiority of our hash based data structure.
We first apply the same scale of BCC box to MISA-MD and LAMMPS,
and monitor the average memory usage of both programs.
In case 1 and case 2 in \cref{tab:memory_usage_of_md}, simulation box of $600 \times 600 \times 600$ containing $4.32 \times 10^8$ atoms is \change{applied},
and 120 MPI processors is involved for both MISA-MD and LAMMPS.
The results of case 1 and case 2 show that the memory usage of LAMMPS is 2.7 times larger than MISA-MD.
In the second test case set, as shown in case 3 and case 4 in \cref{tab:memory_usage_of_md},
we use the full node memory for both MISA-MD and LAMMPS simulation, 
and \change{record the maximum} simulation box they can reach. 
The results show that, given the same memory resource, MISA-MD can simulate much larger scale than LAMMPS.
\change{In more detail}, the number of atoms MISA-MD can simulate is 4.39 times than LAMMPS can reach.

\begin{table}[h]
    \caption{Memory Usage of MISA-MD Compared with LAMMPS on TianHe-2}
    \begin{threeparttable}[t]
    \centering
    \begin{tabular}{llcccc}
        \toprule
         No.    & Program  & Cores & Box & Mem usage & Mem per atom \\
                &          &       &         & (GiB/core)& (Bytes/atom) \\
        \midrule
        1   & MISA-MD & 120 & $600^3$  & 0.64 & 191.2 \\ 
        2   & LAMMPS  & 120 & $600^3$  & 1.77 & 527.2 \\ 
        3   & MISA-MD & 120 & $1094^3$ & 2.66 & 130.9 \\ 
        4   & LAMMPS  & 120 & $668^3$  & 2.66 & 574.9 \\ 
        \bottomrule
    \end{tabular}
   \end{threeparttable}
    \label{tab:memory_usage_of_md}
\end{table}
For computing efficiency, LAMMPS uses neighbor lists (Verlet lists) to keep track of nearby particles.
In the above test cases, the average number of each atom’s neighbor is 56 in LAMMPS.
Thus, it would takes $56 \times 4 = 224$ bytes to index neighbor atoms for each atom,
in which each neighbor index takes 4 bytes.
In test case 4 in \cref{tab:memory_usage_of_md}, the neighbor index memory occupies $38.86\%$ of the total memory.
While in MISA-MD, each atom only occupy 104 bytes for storing basic atoms information (such as velocity, electron charge density),
and neighbor atoms is indexed by index offsets which only consume several KiB memory. 
Therefore, the new hash based data structure in MISA-MD can efficiently reduce the memory usage
and can expand MD simulation to larger scale.

\subsubsection{Computational Performance}
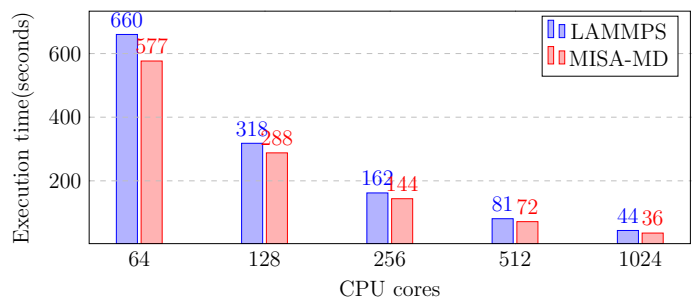
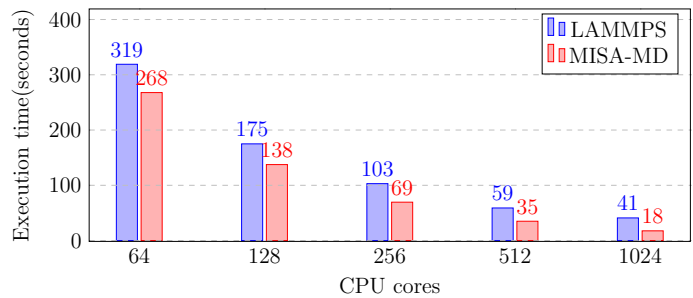
\begin{figure}
    \centering
    \begin{subfigure}[]{0.49\textwidth}
        \begin{adjustbox}{scale=0.70,center}
             
\begin{tikzpicture}
\begin{axis}[
        grid style=dashed,
        ymajorgrids=true, 
        x tick label style={/pgf/number format/1000 sep=},
        ylabel=Execution time(seconds),
        xlabel=CPU cores,
        xtick=data,
        axis on top,
        enlargelimits=0.05,
        ymax=700,
        ybar,
        bar width=0.4cm,
        height=6cm, width=13cm,
        enlarge x limits={value=0.1,upper},
        enlarge x limits=true,
        tickwidth=0pt,
        symbolic x coords={64,128,256,512,1024,},
        nodes near coords={
                \pgfmathprintnumber[precision=0]{\pgfplotspointmeta}
        }
        ]
\addplot 
        coordinates {
                (64,660)
                (128,318)
                (256,162)
                (512,81)
                (1024,44)
        };

\addplot 
        coordinates {
                (64,  576.55)
                (128, 287.90)
                (256, 143.84)
                (512, 71.76)
                (1024, 35.87)
        };



\legend{LAMMPS, MISA-MD}
\end{axis}

\end{tikzpicture}
        \end{adjustbox}
        \caption{Computational performance of MISA-MD and LAMMPS on Tianhe 2.}
        \label{fig:strong_scalability}
    \end{subfigure}
    \qquad
    \begin{subfigure}[]{0.49\textwidth}
        \begin{adjustbox}{scale=0.70,center}
             
\begin{tikzpicture}
\begin{axis}[
        grid style=dashed,
        ymajorgrids=true, 
        x tick label style={/pgf/number format/1000 sep=},
        ylabel=Execution time(seconds),
        xlabel=CPU cores,
        xtick=data,
        axis on top,
        enlargelimits=0.05,
        ymax=400,
        ybar,
        bar width=0.4cm,
        height=6cm, width=13cm,
        enlarge x limits={value=0.1,upper},
        enlarge x limits=true,
        tickwidth=0pt,
        symbolic x coords={64,128,256,512,1024,},
        nodes near coords={
                \pgfmathprintnumber[precision=0]{\pgfplotspointmeta}
        }
        ]

\addplot
        coordinates {
                (64, 319)
                (128, 175)
                (256, 103)
                (512, 59)
                (1024, 41) 
};

\addplot
        coordinates {
                (64, 267.78)
                (128, 137.51)
                (256, 69.48)
                (512, 35.01)
                (1024, 17.63)
};


        
\legend{LAMMPS, MISA-MD}
\end{axis}

\end{tikzpicture}
        \end{adjustbox}
        \caption{Computational performance of MISA-MD and LAMMPS on Sunway Taihulight.
        Acceleration of MISA-MD using CPEs is disabled for principle of fairness.}
        \label{fig:strong_scalability}
    \end{subfigure}

    \caption{Computational performance comparison of MISA-MD and LAMMPS on Sunway Taihulight and Intel platforms
    with $x$ axis logarithmic scaled.
    Simulation time values are annotated on the top of bars.}
    \label{fig:md_perf_compare}
\end{figure}

The execution time is another impact to expand the scale of MD simulation.
We expect simulation program can execute as fast as it could,
thus it can simulate more time steps in the same spatial scale or
finish as soon as possible in the same temporal and spatial scale under the given computing resources.

To compare the performance between MISA-MD and LAMMPS,
test cases with fixed number of atoms are designed for both MISA-MD and LAMMPS simulation.
We start our tests for MISA-MD and LAMMPS on Sunway-Taihulight and TianHe-2 supercomputer.
\new{
On Sunway-TaihuLight, both MISA-MD and LAMMPS were compiled using ``-O2'' optimization flag by sw5gcc/sw5g++(a customized C/C++ compiler based on gcc version 5.3.0).
It is worth noting that, CPEs acceleration of MISA-MD is disabled for principle of fairness because of the lack of EAM calculation acceleration on CPEs of LAMMPS.
On TianHe-2, both the programs were also compiled using ``-O2'' optimization flag by the same compiler(gcc 5.3.0).
}

By giving varies of atoms number and involving 64, 128, 256, 512 and 1024 CPU cores respectively for both MISA-MD and LAMMPS,
and recording execution time of both programs, the results of computational performance are obtained.
The test cases on Tianhe-2 supercomputer use a $512 \times 512 \times 512$ box containing $2.68 \times 10^8$ atoms for varies of CPU cores.
While for test cases on Sunway-Taihulight, simulation box of $256 \times 256 \times 256$ containing $3.36 \times 10^7$ atoms is introduced.

The performance results for both MISA-MD code and LAMMPS code on \change{Intel} and sunway platform are shown in \cref{fig:md_perf_compare}.
\new{On Sunway platform, compared with LAMMPS package,
computational performance of MISA-MD is increased by 19.02\% to 132.56\% on Sunway-Taihulight.
On \change{Intel} platform of TianHe 2, the performance increment of MISA-MD is from 10.46\% to 22.67\% compared with LAMMPS packages.
}
Results indicate that, \textbf{
    MISA-MD has a better performance than LAMMPS for all performance test cases above.
 }

In LAMMPS, the neighbor list is updated every few timesteps and its cost can be quite expensive.
While in MISA-MD, hash index can be updated in each timestep,
but the time consuming of hash index updating is much less than neighbor list updating in LAMMPS
if the system does not change very frequently.
In cascade collision simulation of metal materials, position of most atoms does not change very frequently.
Thus, in hash index updating stage, most atoms' hash index does not need to be updated,
which can lead a better performance of MISA-MD than LAMMPS.
We should also point out that simulation of other types of crystal material can also be realistic and operable,
and can achieve a quite respectable performance.

\subsection{Scalability}
In this subsection, we will discuss our study on the scalability of our MISA-MD code on Sunway TaihuLight supercomputer.
We use cascade collision cases of pure $\alpha-$Fe system at temperature of 600 K as the case for the scalability tests.

For the strong scalability, a simulation box with 655 billion ($6.55\times 10^{11}$) atoms is presented
and take the performance of \SI{520000}{} cores (8000 core groups) as baseline.
\cref{fig:strong_scalability} shows the parallel efficiency of this case for strong scalability
when the number of cores(including CPE cores) varies from \SI{520000}{} to \SI{8320000}{}.
We can found that when the number of atoms per core group drops to 1/16 of the number,
the parallel efficiency drops from 100\% to 79.55\%, which is quite great for the parallel efficiency.

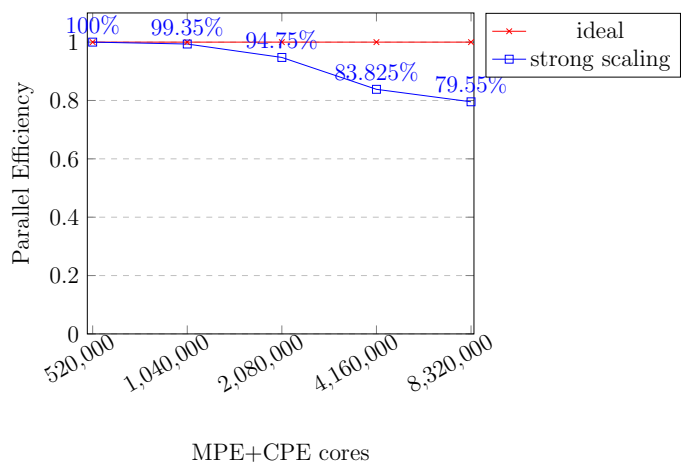
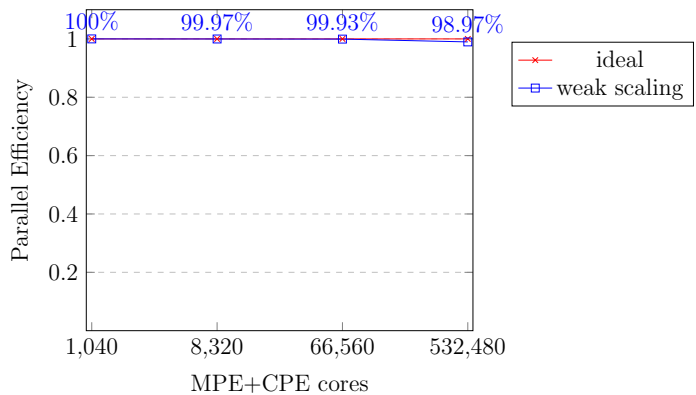
\begin{figure}
    \centering
    \begin{subfigure}[]{0.49\textwidth}
        \begin{adjustbox}{scale=0.75,center}
             
\begin{tikzpicture}
    \begin{axis}[
        xlabel={MPE+CPE cores},
        ylabel={ Parallel Efficiency },
        xmin=500000, xmax=8500000,
        ymin=0, ymax=1.1,
        xmode=log,
        log basis x={2},
        log ticks with fixed point,
        x tick label style={rotate=30, anchor=east, xshift=0.4cm, yshift=-0.3cm},
        xlabel style={yshift=-0.5cm}, 
        xtick={520000, 1040000, 2080000, 4160000, 8320000},
        ytick={0, 0.2, 0.4, 0.6, 0.8, 1.0},
        legend pos=outer north east,
        ymajorgrids=true,
        grid style=dashed,
    ]

    \addplot[
        color=red,
        mark=x,
        ]
        coordinates {
        (520000, 1)
        (1040000, 1)
        (2080000, 1)
        (4160000, 1)
        (8320000, 1)
    };

    \addplot[
        color=blue,
        mark=square,
        nodes near coords,
        point meta=explicit symbolic,
        visualization depends on=\thisrow{alignment} \as \alignment,
        ]
        table [
        meta index=2 
        ] {
            x       y       label       alignment
            520000   1         100\%       -40
            1040000   0.9935   99.35\%     -40
            2080000   0.9475   94.75\%     160
            4160000   0.83825  83.825\%    -40
            8320000   0.7955   79.55\%     -40
        };
        \addlegendentry{ ideal }
        \addlegendentry{ strong scaling}
    \end{axis}

\end{tikzpicture}
        \end{adjustbox}
        \caption{Strong scalability for $6.55 \times 10^{11}$ atoms with $x$ axis logarithmic scaled.
        The number above the line is the parallel efficiency.}
        \label{fig:strong_scalability}
    \end{subfigure}
    \qquad
    \begin{subfigure}[]{0.49\textwidth}
        \begin{adjustbox}{scale=0.75,center}
            \begin{tikzpicture}
    \pgfplotstableread[col sep=space]{
        x
        1,040
        8,320
        66,560
        532,480
    }\realtable

    \begin{axis}[
        xlabel={MPE+CPE cores},
        ylabel={ Parallel Efficiency },
        xmin=950, xmax=580000,
        ymin=0.0, ymax=1.1,
        xmode=log,
        log basis x={2},
        log ticks with fixed point,
        xtick=data,
        xticklabels from table={\realtable}{x},
        ytick={0.2, 0.4, 0.6, 0.8, 1.0},
        legend pos=outer north east,
        legend style={at={(1.1,0.9)},anchor=north west},
        ymajorgrids=true,
        grid style=dashed,
    ]

    \addplot[
        color=red,
        mark=x,
        ]
        coordinates {
        (1040, 1)
        (8320, 1)
        (66560, 1)
        (532480, 1)
    };
    \addplot[
        color=blue,
        mark=square,
        nodes near coords,
        point meta=explicit symbolic,
        visualization depends on=\thisrow{alignment} \as \alignment,
        ]
        table [
        meta index=2 
        ] {
            x       y       label       alignment
            1040   1         100\%      -40
            8320   0.9997    99.97\%     -40
            66560  0.9993    99.93\%    160
            532480 0.9897    98.97\%    -40
        };
        \addlegendentry{ ideal }
        \addlegendentry{ weak scaling}
    \end{axis}
\end{tikzpicture}
        \end{adjustbox}
        \caption{Weak scalability for $8.39 \times 10^{6}$ atoms per core group (or MPI process) with $x$ axis logarithmic scaled. 
        The number above the line is the parallel efficiency.}
        \label{fig:weak_scalability}
    \end{subfigure}
    \caption{Results of strong scalability and weak scalability performance tests.}
    \label{fig:md_performance}
\end{figure}

The weak scalability performance is shown in \cref{fig:weak_scalability}.
In the weak scalability tests, we initialize the simulation system with $8.39 \times 10^6$ atoms per core group (MPI process) 
and take the performance of 1040 cores (16 core groups) as baseline.
It can be seen that there is almost no performance loss as the scale increase.
The parallel efficiency only goes down to 98.97\% when the number of cores is increased to 512 times of the baseline.

Results from both strong scalability and weak scalability indicate that 
our MISA-MD code have the ability to expand MD simulation to larger scale with large scale parallelism.
Moreover, we also achieve $3.02 \times 10^{13}$ atoms cascade collision simulation at \SI{131072}{} core groups on Sunway TaihuLight supercomputer.
To our best knowledge, it holds the world record of the largest molecular dynamics simulation currently.

\section{Conclusion}
In this work, a new molecular dynamics code, MISA-MD,
is developed for simulating material evolution in cascade collision.
As a basic innovation,
a new hash based data structure is proposed for efficient atoms storage and quick neighbor indexing.
To fully utilize the performance of SW26010 processor,
the EAM potential interaction computation is also implementation on Sunway TaihuLight
with some efficient acceleration and optimization strategies,
including a coloring method for avoiding writing conflict on sunway CPEs, double-buffer DMA and data reuse strategies etc.
With the new data structure and accelerating of EAM interaction computation on SW26010 processor,
we can extend MD simulation to larger temporal scale and longer spatial scales.
It is worth noting that, many of our strategies are general and could also be implemented 
in other multi-core processors and heterogeneous platforms.
For example, the coloring method for sunway CPEs can be easily migrated 
to OpenMP multiple threads program model or CUDA platforms.
With these methods above,
performance of 79.5\% parallel efficiency is obtained in strong scaling simulation on Sunway TaihuLight system,
and simulation of more than thirty-trillion atoms is performed using \SI{131072}{} sunway core groups.
Moreover, experiments show that MISA-MD provides high accuracy of cascade collision simulation
and has less memory overhead and higher performance compared with classical molecular dynamics code LAMMPS.

Moreover, MISA-MD is designed to couple its simulation result with kinetic Monte Carlo (kMC) method\cite{sickafus_introduction_2007}
and Cluster Dynamics (CD) method\cite{terrier_cluster_2017} for further multisale evolution simulation
\cite{caturla_multiscale_2001,molnar_multiscale_2012,wirth_multiscale_2004,fu_multiscale_2004} 
of cluster and defects in long temporal scale(seconds to years).

\section*{Acknowledgment}


This article is the results of the research project funded
by The National Natural Science Foundation of China under Grant U1867217.





\bibliographystyle{elsarticle/elsarticle-num}
\bibliography{ref.bib}

\begin{thebibliography}{10}
\expandafter\ifx\csname url\endcsname\relax
  \def\url#1{\texttt{#1}}\fi
\expandafter\ifx\csname urlprefix\endcsname\relax\def\urlprefix{URL }\fi
\expandafter\ifx\csname href\endcsname\relax
  \def\href#1#2{#2} \def\path#1{#1}\fi

\bibitem{stoller_primary_2012}
R.~Stoller,
  \href{https://linkinghub.elsevier.com/retrieve/pii/B9780080560335000276}{Primary
  {Radiation} {Damage} {Formation}}, in: Comprehensive {Nuclear} {Materials},
  Elsevier, 2012, pp. 293--332.
\newblock \href {http://dx.doi.org/10.1016/B978-0-08-056033-5.00027-6}
  {\path{doi:10.1016/B978-0-08-056033-5.00027-6}}.
\newline\urlprefix\url{https://linkinghub.elsevier.com/retrieve/pii/B9780080560335000276}

\bibitem{knaster_materials_2016}
J.~Knaster, A.~Moeslang, T.~Muroga,
  \href{http://www.nature.com/doifinder/10.1038/nphys3735}{Materials research
  for fusion}, Nature Physics (2016) 424--434\href
  {http://dx.doi.org/10.1038/nphys3735} {\path{doi:10.1038/nphys3735}}.
\newline\urlprefix\url{http://www.nature.com/doifinder/10.1038/nphys3735}

\bibitem{nordlund_primary_2018}
K.~Nordlund, S.~J. Zinkle, A.~E. Sand, F.~Granberg, R.~S. Averback, R.~E.
  Stoller, T.~Suzudo, L.~Malerba, F.~Banhart, W.~J. Weber, F.~Willaime, S.~L.
  Dudarev, D.~Simeone,
  \href{https://linkinghub.elsevier.com/retrieve/pii/S002231151831016X}{Primary
  radiation damage: {A} review of current understanding and models}, Journal of
  Nuclear Materials 512 (2018) 450--479.
\newblock \href {http://dx.doi.org/10.1016/j.jnucmat.2018.10.027}
  {\path{doi:10.1016/j.jnucmat.2018.10.027}}.
\newline\urlprefix\url{https://linkinghub.elsevier.com/retrieve/pii/S002231151831016X}

\bibitem{nordlund_historical_2019}
K.~Nordlund,
  \href{https://linkinghub.elsevier.com/retrieve/pii/S0022311518314703}{Historical
  review of computer simulation of radiation effects in materials}, Journal of
  Nuclear Materials 520 (2019) 273--295, 00000.
\newblock \href {http://dx.doi.org/10.1016/j.jnucmat.2019.04.028}
  {\path{doi:10.1016/j.jnucmat.2019.04.028}}.
\newline\urlprefix\url{https://linkinghub.elsevier.com/retrieve/pii/S0022311518314703}

\bibitem{peng_shockwave_2018}
Q.~Peng, F.~Meng, Y.~Yang, C.~Lu, H.~Deng, L.~Wang, S.~De, F.~Gao,
  \href{http://www.nature.com/articles/s41467-018-07102-3}{Shockwave generates
  {\textless} 100 {\textgreater} dislocation loops in bcc iron}, Nature
  Communications 9~(1), 00000.
\newblock \href {http://dx.doi.org/10.1038/s41467-018-07102-3}
  {\path{doi:10.1038/s41467-018-07102-3}}.
\newline\urlprefix\url{http://www.nature.com/articles/s41467-018-07102-3}

\bibitem{fu_molecular_2019}
J.~Fu, Y.~Chen, J.~Fang, N.~Gao, W.~Hu, C.~Jiang, H.-B. Zhou, G.-H. Lu, F.~Gao,
  H.~Deng,
  \href{https://linkinghub.elsevier.com/retrieve/pii/S0022311519304416}{Molecular
  dynamics simulations of high-energy radiation damage in {W} and {W}–{Re}
  alloys}, Journal of Nuclear Materials 524 (2019) 9--20, 00000.
\newblock \href {http://dx.doi.org/10.1016/j.jnucmat.2019.06.027}
  {\path{doi:10.1016/j.jnucmat.2019.06.027}}.
\newline\urlprefix\url{https://linkinghub.elsevier.com/retrieve/pii/S0022311519304416}

\bibitem{plimpton1993fast}
S.~Plimpton, Fast parallel algorithms for short-range molecular dynamics, Tech.
  rep., Sandia National Labs., Albuquerque, NM (United States) (1993).

\bibitem{duan_redesigning_2018}
X.~Duan, P.~Gao, T.~Zhang, M.~Zhang, W.~Liu, W.~Zhang, W.~Xue, H.~Fu, L.~Gan,
  D.~Chen, X.~Meng, G.~Yang,
  \href{https://ieeexplore.ieee.org/document/8665774/}{Redesigning {LAMMPS} for
  {Peta}-{Scale} and {Hundred}-{Billion}-{Atom} {Simulation} on {Sunway}
  {TaihuLight}}, in: {SC18}: {International} {Conference} for {High}
  {Performance} {Computing}, {Networking}, {Storage} and {Analysis}, IEEE,
  Dallas, TX, USA, 2018, pp. 148--159, 00000.
\newblock \href {http://dx.doi.org/10.1109/SC.2018.00015}
  {\path{doi:10.1109/SC.2018.00015}}.
\newline\urlprefix\url{https://ieeexplore.ieee.org/document/8665774/}

\bibitem{niethammer_ls1_2014}
C.~Niethammer, S.~Becker, M.~Bernreuther, M.~Buchholz, W.~Eckhardt,
  A.~Heinecke, S.~Werth, H.-J. Bungartz, C.~W. Glass, H.~Hasse, J.~Vrabec,
  M.~Horsch, \href{https://pubs.acs.org/doi/10.1021/ct500169q}{\textit{ls1
  mardyn} : {The} {Massively} {Parallel} {Molecular} {Dynamics} {Code} for
  {Large} {Systems}}, Journal of Chemical Theory and Computation 10~(10) (2014)
  4455--4464.
\newblock \href {http://dx.doi.org/10.1021/ct500169q}
  {\path{doi:10.1021/ct500169q}}.
\newline\urlprefix\url{https://pubs.acs.org/doi/10.1021/ct500169q}

\bibitem{tersoff_new_1988}
J.~Tersoff, \href{https://link.aps.org/doi/10.1103/PhysRevB.37.6991}{New
  empirical approach for the structure and energy of covalent systems}, Phys.
  Rev. B 37~(12) (1988) 6991--7000, publisher: American Physical Society.
\newblock \href {http://dx.doi.org/10.1103/PhysRevB.37.6991}
  {\path{doi:10.1103/PhysRevB.37.6991}}.
\newline\urlprefix\url{https://link.aps.org/doi/10.1103/PhysRevB.37.6991}

\bibitem{lennard1931cohesion}
J.~E. Lennard-Jones, Cohesion, Proceedings of the Physical Society 43~(5)
  (1931) 461.

\bibitem{daw_embedded-atom_1984}
M.~S. Daw, M.~I. Baskes,
  \href{https://link.aps.org/doi/10.1103/PhysRevB.29.6443}{Embedded-atom
  method: {Derivation} and application to impurities, surfaces, and other
  defects in metals}, Physical Review B 29~(12) (1984) 6443--6453, 00000.
\newblock \href {http://dx.doi.org/10.1103/PhysRevB.29.6443}
  {\path{doi:10.1103/PhysRevB.29.6443}}.
\newline\urlprefix\url{https://link.aps.org/doi/10.1103/PhysRevB.29.6443}

\bibitem{rapaport_art_2004}
D.~C. Rapaport, The art of molecular dynamics simulation, 2nd Edition,
  Cambridge University Press, Cambridge, UK ; New York, NY, 2004.

\bibitem{law_algorithm_2019}
T.~Law, J.~Hancox, S.~Wright, S.~Jarvis,
  \href{https://linkinghub.elsevier.com/retrieve/pii/S0743731519302047}{An
  algorithm for computing short-range forces in molecular dynamics simulations
  with non-uniform particle densities}, Journal of Parallel and Distributed
  Computing 130 (2019) 1--11.
\newblock \href {http://dx.doi.org/10.1016/j.jpdc.2019.03.008}
  {\path{doi:10.1016/j.jpdc.2019.03.008}}.
\newline\urlprefix\url{https://linkinghub.elsevier.com/retrieve/pii/S0743731519302047}

\bibitem{plimpton_parallel_1992}
S.~J. Plimpton, B.~A. Hendrickson,
  \href{http://link.springer.com/10.1557/PROC-291-37}{Parallel molecular
  dynamics with the embedded atom method} 291  37.
\newblock \href {http://dx.doi.org/10.1557/PROC-291-37}
  {\path{doi:10.1557/PROC-291-37}}.
\newline\urlprefix\url{http://link.springer.com/10.1557/PROC-291-37}

\bibitem{zhigilei2014introduction}
L.~Zhigilei, Introduction to atomistic simulations, University of Virginia),
  MSE 4270 (2014) 6270.

\bibitem{fu_sunway_2016}
H.~Fu, J.~Liao, J.~Yang, L.~Wang, Z.~Song, X.~Huang, C.~Yang, W.~Xue, F.~Liu,
  F.~Qiao, W.~Zhao, X.~Yin, C.~Hou, C.~Zhang, W.~Ge, J.~Zhang, Y.~Wang,
  C.~Zhou, G.~Yang,
  \href{http://link.springer.com/10.1007/s11432-016-5588-7}{The {Sunway}
  {TaihuLight} supercomputer: system and applications}, Science China
  Information Sciences 59~(7).
\newblock \href {http://dx.doi.org/10.1007/s11432-016-5588-7}
  {\path{doi:10.1007/s11432-016-5588-7}}.
\newline\urlprefix\url{http://link.springer.com/10.1007/s11432-016-5588-7}

\bibitem{yang_application_2018}
G.-W. Yang, H.-H. Fu,
  \href{http://link.springer.com/10.1631/FITEE.1800459}{Application software
  beyond exascale: challenges and possible trends}, Frontiers of Information
  Technology \& Electronic Engineering 19~(10) (2018) 1267--1272.
\newblock \href {http://dx.doi.org/10.1631/FITEE.1800459}
  {\path{doi:10.1631/FITEE.1800459}}.
\newline\urlprefix\url{http://link.springer.com/10.1631/FITEE.1800459}

\bibitem{noauthor_homegrown_2015}
\href{http://engine.scichina.com/doi/10.1360/N112014-00299}{A homegrown
  many-core processor architecture for high-performance computing}, SCIENTIA
  SINICA Informationis00006.
\newblock \href {http://dx.doi.org/10.1360/N112014-00299}
  {\path{doi:10.1360/N112014-00299}}.
\newline\urlprefix\url{http://engine.scichina.com/doi/10.1360/N112014-00299}

\bibitem{berendsen1995gromacs}
H.~J. Berendsen, D.~van~der Spoel, R.~van Drunen, Gromacs: a message-passing
  parallel molecular dynamics implementation, Computer physics communications
  91~(1-3) (1995) 43--56.

\bibitem{zhang_sw_gromacs_2019}
T.~Zhang, L.~Gan, H.~Fu, W.~Xue, W.~Liu, G.~Yang, Y.~Li, P.~Gao, Q.~Shao,
  M.~Shao, M.~Zhang, J.~Zhang, X.~Duan, Z.~Liu,
  \href{http://dl.acm.org/citation.cfm?doid=3295500.3356190}{{SW}\_gromacs:
  accelerate {GROMACS} on {Sunway} {TaihuLight}}, in: Proceedings of the
  {International} {Conference} for {High} {Performance} {Computing},
  {Networking}, {Storage} and {Analysis} on - {SC} '19, ACM Press, Denver,
  Colorado, 2019, pp. 1--14.
\newblock \href {http://dx.doi.org/10.1145/3295500.3356190}
  {\path{doi:10.1145/3295500.3356190}}.
\newline\urlprefix\url{http://dl.acm.org/citation.cfm?doid=3295500.3356190}

\bibitem{phillips_namd_2002}
J.~Phillips, {Gengbin Zheng}, S.~Kumar, L.~Kale,
  \href{http://ieeexplore.ieee.org/document/1592872/}{{NAMD}: {Biomolecular}
  {Simulation} on {Thousands} of {Processors}}, in: {ACM}/{IEEE} {SC} 2002
  {Conference} ({SC}'02), IEEE, Baltimore, MD, USA, 2002, pp. 36--36.
\newblock \href {http://dx.doi.org/10.1109/SC.2002.10019}
  {\path{doi:10.1109/SC.2002.10019}}.
\newline\urlprefix\url{http://ieeexplore.ieee.org/document/1592872/}

\bibitem{kale_charm_1993}
L.~V. Kale, S.~Krishnan,
  \href{http://portal.acm.org/citation.cfm?doid=167962.165874}{{CHARM}++: a
  portable concurrent object oriented system based on {C}++}, ACM SIGPLAN
  Notices 28~(10) (1993) 91--108.
\newblock \href {http://dx.doi.org/10.1145/167962.165874}
  {\path{doi:10.1145/167962.165874}}.
\newline\urlprefix\url{http://portal.acm.org/citation.cfm?doid=167962.165874}

\bibitem{tchipev_twetris:_2019}
N.~Tchipev, S.~Seckler, M.~Heinen, J.~Vrabec, F.~Gratl, M.~Horsch,
  M.~Bernreuther, C.~W. Glass, C.~Niethammer, N.~Hammer, B.~Krischok, M.~Resch,
  D.~Kranzlmüller, H.~Hasse, H.-J. Bungartz, P.~Neumann,
  \href{http://journals.sagepub.com/doi/10.1177/1094342018819741}{{TweTriS}:
  {Twenty} trillion-atom simulation}, The International Journal of High
  Performance Computing Applications (2019) 109434201881974\href
  {http://dx.doi.org/10.1177/1094342018819741}
  {\path{doi:10.1177/1094342018819741}}.
\newline\urlprefix\url{http://journals.sagepub.com/doi/10.1177/1094342018819741}

\bibitem{hu_crystal_2017}
C.~Hu, H.~Bai, X.~He, B.~Zhang, N.~Nie, X.~Wang, Y.~Ren,
  \href{https://linkinghub.elsevier.com/retrieve/pii/S001046551630193X}{Crystal
  {MD}: {The} massively parallel molecular dynamics software for metal with
  {BCC} structure}, Computer Physics Communications 211 (2017) 73--78.
\newblock \href {http://dx.doi.org/10.1016/j.cpc.2016.07.011}
  {\path{doi:10.1016/j.cpc.2016.07.011}}.
\newline\urlprefix\url{https://linkinghub.elsevier.com/retrieve/pii/S001046551630193X}

\bibitem{hu_kernel_2017}
C.~Hu, X.~Wang, J.~Li, X.~He, S.~Li, Y.~Feng, S.~Yang, H.~Bai,
  \href{https://linkinghub.elsevier.com/retrieve/pii/S0010465516301928}{Kernel
  optimization for short-range molecular dynamics}, Computer Physics
  Communications 211 (2017) 31--40.
\newblock \href {http://dx.doi.org/10.1016/j.cpc.2016.07.010}
  {\path{doi:10.1016/j.cpc.2016.07.010}}.
\newline\urlprefix\url{https://linkinghub.elsevier.com/retrieve/pii/S0010465516301928}

\bibitem{germann_25_nodate}
T.~C. Germann, K.~Kadau, P.~S. Lomdahl, {25} {Tflop}/{s} {Multibillion}-{Atom}
  {Molecular} {Dynamics} {Simulations} and {Visualization} {Analysis} on
  {BlueGene}/{L}  13.

\bibitem{verlet_computer_1967}
L.~Verlet, \href{https://link.aps.org/doi/10.1103/PhysRev.159.98}{Computer
  "{Experiments}" on {Classical} {Fluids}. {I}. {Thermodynamical} {Properties}
  of {Lennard}-{Jones} {Molecules}}, Phys. Rev. 159~(1) (1967) 98--103,
  publisher: American Physical Society.
\newblock \href {http://dx.doi.org/10.1103/PhysRev.159.98}
  {\path{doi:10.1103/PhysRev.159.98}}.
\newline\urlprefix\url{https://link.aps.org/doi/10.1103/PhysRev.159.98}

\bibitem{hockney_quiet_1974}
R.~W. Hockney, S.~P. Goel, J.~W. Eastwood,
  \href{http://www.sciencedirect.com/science/article/pii/0021999174900102}{Quiet
  high-resolution computer models of a plasma}, Journal of Computational
  Physics 14~(2) (1974) 148 -- 158.
\newblock \href
  {http://dx.doi.org/https://doi.org/10.1016/0021-9991(74)90010-2}
  {\path{doi:https://doi.org/10.1016/0021-9991(74)90010-2}}.
\newline\urlprefix\url{http://www.sciencedirect.com/science/article/pii/0021999174900102}

\bibitem{lin_evaluating_2018}
J.~Lin, Z.~Xu, L.~Cai, A.~Nukada, S.~Matsuoka,
  \href{https://linkinghub.elsevier.com/retrieve/pii/S0167819118301820}{Evaluating
  the {SW26010} many-core processor with a micro-benchmark suite for
  performance optimizations}, Parallel Computing 77 (2018) 128--143.
\newblock \href {http://dx.doi.org/10.1016/j.parco.2018.06.001}
  {\path{doi:10.1016/j.parco.2018.06.001}}.
\newline\urlprefix\url{https://linkinghub.elsevier.com/retrieve/pii/S0167819118301820}

\bibitem{bonny_ternary_2009}
G.~Bonny, R.~Pasianot, N.~Castin, L.~Malerba,
  \href{http://www.tandfonline.com/doi/abs/10.1080/14786430903299824}{Ternary
  {Fe}–{Cu}–{Ni} many-body potential to model reactor pressure vessel
  steels: {First} validation by simulated thermal annealing}, Philosophical
  Magazine 89~(34-36) (2009) 3531--3546, 00000.
\newblock \href {http://dx.doi.org/10.1080/14786430903299824}
  {\path{doi:10.1080/14786430903299824}}.
\newline\urlprefix\url{http://www.tandfonline.com/doi/abs/10.1080/14786430903299824}

\bibitem{li_massively_2018}
S.~Li, B.~Wu, Y.~Zhang, X.~Wang, J.~Li, C.~Hu, J.~Wang, Y.~Feng, N.~Nie,
  \href{http://dl.acm.org/citation.cfm?doid=3225058.3225064}{Massively
  {Scaling} the {Metal} {Microscopic} {Damage} {Simulation} on {Sunway}
  {TaihuLight} {Supercomputer}}, in: Proceedings of the 47th {International}
  {Conference} on {Parallel} {Processing} - {ICPP} 2018, ACM Press, Eugene, OR,
  USA, 2018, pp. 1--11.
\newblock \href {http://dx.doi.org/10.1145/3225058.3225064}
  {\path{doi:10.1145/3225058.3225064}}.
\newline\urlprefix\url{http://dl.acm.org/citation.cfm?doid=3225058.3225064}

\bibitem{cai_openacc_2018}
L.~Cai, Y.-C. Wang, W.~Tang, B.~Wang, S.~Ethier, Z.~Liu, J.~Lin,
  \href{https://ieeexplore.ieee.org/document/8514862/}{{OpenACC} vs the native
  programming on sunway {TaihuLight}: A case study with {GTC}-p}, in: 2018
  {IEEE} International Conference on Cluster Computing ({CLUSTER}), {IEEE}, pp.
  88--97.
\newblock \href {http://dx.doi.org/10.1109/CLUSTER.2018.00021}
  {\path{doi:10.1109/CLUSTER.2018.00021}}.
\newline\urlprefix\url{https://ieeexplore.ieee.org/document/8514862/}

\bibitem{liu_efficient_2011}
Y.~Liu, C.~Hu, C.~Zhao,
  \href{https://linkinghub.elsevier.com/retrieve/pii/S0010465511000270}{Efficient
  parallel implementation of {Ewald} summation in molecular dynamics
  simulations on multi-core platforms}, Computer Physics Communications 182~(5)
  (2011) 1111--1119.
\newblock \href {http://dx.doi.org/10.1016/j.cpc.2011.01.007}
  {\path{doi:10.1016/j.cpc.2011.01.007}}.
\newline\urlprefix\url{https://linkinghub.elsevier.com/retrieve/pii/S0010465511000270}

\bibitem{hu_efficient_2009}
C.~Hu, Y.~Liu, J.~Li,
  \href{http://ieeexplore.ieee.org/document/5363184/}{Efficient {Parallel}
  {Implementation} of {Molecular} {Dynamics} with {Embedded} {Atom} {Method} on
  {Multi}-core {Platforms}}, in: 2009 {International} {Conference} on
  {Parallel} {Processing} {Workshops}, IEEE, Vienna, Austria, 2009, pp.
  121--129.
\newblock \href {http://dx.doi.org/10.1109/ICPPW.2009.24}
  {\path{doi:10.1109/ICPPW.2009.24}}.
\newline\urlprefix\url{http://ieeexplore.ieee.org/document/5363184/}

\bibitem{norgett_proposed_1975}
M.~Norgett, M.~Robinson, I.~Torrens,
  \href{https://linkinghub.elsevier.com/retrieve/pii/0029549375900357}{A
  proposed method of calculating displacement dose rates}, Nuclear Engineering
  and Design 33~(1) (1975) 50--54.
\newblock \href {http://dx.doi.org/10.1016/0029-5493(75)90035-7}
  {\path{doi:10.1016/0029-5493(75)90035-7}}.
\newline\urlprefix\url{https://linkinghub.elsevier.com/retrieve/pii/0029549375900357}

\bibitem{sickafus_introduction_2007}
A.~F. Voter,
  \href{http://link.springer.com/10.1007/978-1-4020-5295-8_1}{{INTRODUCTION}
  {TO} {THE} {KINETIC} {MONTE} {CARLO} {METHOD}}, in: K.~E. Sickafus, E.~A.
  Kotomin, B.~P. Uberuaga (Eds.), Radiation {Effects} in {Solids}, Vol. 235,
  Springer Netherlands, Dordrecht, 2007, pp. 1--23.
\newblock \href {http://dx.doi.org/10.1007/978-1-4020-5295-8_1}
  {\path{doi:10.1007/978-1-4020-5295-8_1}}.
\newline\urlprefix\url{http://link.springer.com/10.1007/978-1-4020-5295-8_1}

\bibitem{terrier_cluster_2017}
P.~Terrier, M.~Athènes, T.~Jourdan, G.~Adjanor, G.~Stoltz,
  \href{https://linkinghub.elsevier.com/retrieve/pii/S0021999117305867}{Cluster
  dynamics modelling of materials: {A} new hybrid deterministic/stochastic
  coupling approach}, Journal of Computational Physics 350 (2017) 280--295.
\newblock \href {http://dx.doi.org/10.1016/j.jcp.2017.08.015}
  {\path{doi:10.1016/j.jcp.2017.08.015}}.
\newline\urlprefix\url{https://linkinghub.elsevier.com/retrieve/pii/S0021999117305867}

\bibitem{caturla_multiscale_2001}
M.~J. Caturla, R.~K. Corzine, M.~R. James, G.~A. Greene, Multiscale modeling of
  radiation damage: applications to damage production by {GeV} proton
  irradiation of {Cu} and {W}, and pulsed irradiation effects in {Cu} and {Fe},
  Journal of Nuclear Materials (2001) 11.

\bibitem{molnar_multiscale_2012}
D.~Molnar, R.~Mukherjee, A.~Choudhury, A.~Mora, P.~Binkele, M.~Selzer,
  B.~Nestler, S.~Schmauder,
  \href{https://linkinghub.elsevier.com/retrieve/pii/S1359645412006003}{Multiscale
  simulations on the coarsening of {Cu}-rich precipitates in {$\alpha$}-{Fe}
  using kinetic {Monte} {Carlo}, molecular dynamics and phase-field
  simulations}, Acta Materialia 60~(20) (2012) 6961--6971, 00000.
\newblock \href {http://dx.doi.org/10.1016/j.actamat.2012.08.051}
  {\path{doi:10.1016/j.actamat.2012.08.051}}.
\newline\urlprefix\url{https://linkinghub.elsevier.com/retrieve/pii/S1359645412006003}

\bibitem{wirth_multiscale_2004}
B.~Wirth, G.~Odette, J.~Marian, L.~Ventelon, J.~Young-Vandersall,
  L.~Zepeda-Ruiz,
  \href{https://linkinghub.elsevier.com/retrieve/pii/S0022311504001321}{Multiscale
  modeling of radiation damage in {Fe}-based alloys in the fusion environment},
  Journal of Nuclear Materials 329-333 (2004) 103--111, 00000.
\newblock \href {http://dx.doi.org/10.1016/j.jnucmat.2004.04.156}
  {\path{doi:10.1016/j.jnucmat.2004.04.156}}.
\newline\urlprefix\url{https://linkinghub.elsevier.com/retrieve/pii/S0022311504001321}

\bibitem{fu_multiscale_2004}
C.-C. Fu, J.~D. Torre, F.~Willaime, J.-L. Bocquet, A.~Barbu,
  \href{http://www.nature.com/doifinder/10.1038/nmat1286}{Multiscale modelling
  of defect kinetics in irradiated iron}, Nature Materials 4~(1) (2004) 68--74.
\newblock \href {http://dx.doi.org/10.1038/nmat1286}
  {\path{doi:10.1038/nmat1286}}.
\newline\urlprefix\url{http://www.nature.com/doifinder/10.1038/nmat1286}

\end{thebibliography}







\end{document}